\documentclass[superscriptaddress,reprint,amssymb, nobibnotes, aps, pra]{revtex4-2}

\newcommand{\env}[1]{\texttt{#1}}
\DeclareMathAlphabet{\mathpzc}{OT1}{pzc}{m}{it}

\usepackage{graphicx}
\usepackage[caption=false]{subfig}
\usepackage{epsfig}
\usepackage{epstopdf}
\usepackage{amsmath}
\usepackage{ upgreek }
\usepackage{float}
\usepackage{enumerate}
\usepackage{array}
\usepackage{pifont}
\usepackage{comment}
\usepackage[11pt]{moresize}
\usepackage{datetime}
\usepackage{multirow}

\usepackage{color, colortbl}
\definecolor{LightCyan}{rgb}{0.88,1,1}

\usepackage{mathrsfs}
\newcolumntype{C}[1]{>{\centering\let\newline\\\arraybackslash\hspace{0pt}}m{#1}}
\newcolumntype{N}{@{}m{0pt}@{}}

\begin{document}
	
%\preprint{}

\title{Spin-governed topological surfaces and broken spin-momentum locking in a gyromagnetic medium}



\author{Rajarshi Sen}
\affiliation{Department of Electronics and Electrical Communication Engineering, Indian Institute of Technology Kharagpur, Kharagpur, West Bengal, India}
\author{Sarang Pendharker}
\affiliation{Department of Electronics and Electrical Communication Engineering, Indian Institute of Technology Kharagpur, Kharagpur, West Bengal, India}
\email[]{sarang@ece.iitkgp.ac.in}

%\thanks{}


\date{\today}

\begin{abstract}
Topology of isofrequency surfaces plays a crucial role in characterizing the interaction of an electromagnetic wave with a medium. Thus, engineering the topology in complex media is leading to novel applications, ranging from super-resolution microscopy with hyperbolic metamaterials to sub-wavelength waveguiding structures. Here, we investigate the spin-governed nature of isofrequency surfaces in a general gyromagnetic medium. We show that gyrotropy also plays an important role in the topological properties of a medium, along with the anisotropic permeability and permittivity. Even though the topology primarily depends on permeability, gyrotropy can suppress or support the existence of certain topological surfaces. We reveal the connection between the gyrotropy imposed constraints and the photonic spin-profile of the topological surfaces. The spin-profile along the isofrequency surface is locked to the material, resulting in the non-reciprocity and breaking of the spin-momentum locking in the gyromagnetic medium. Further, we show that the conflict between spin-momentum locking and material locked spin leads to asymmetric mode profile and gyrotropy-induced cutoff in guided wave structures. Our work provides important insights into the underlying link between topology, spin, and non-reciprocity in gyrotropic media.

\end{abstract}

%\keywords{}

\maketitle

\section{Introduction}

\begin{figure}
    \centering
    \includegraphics[width=0.7\linewidth]{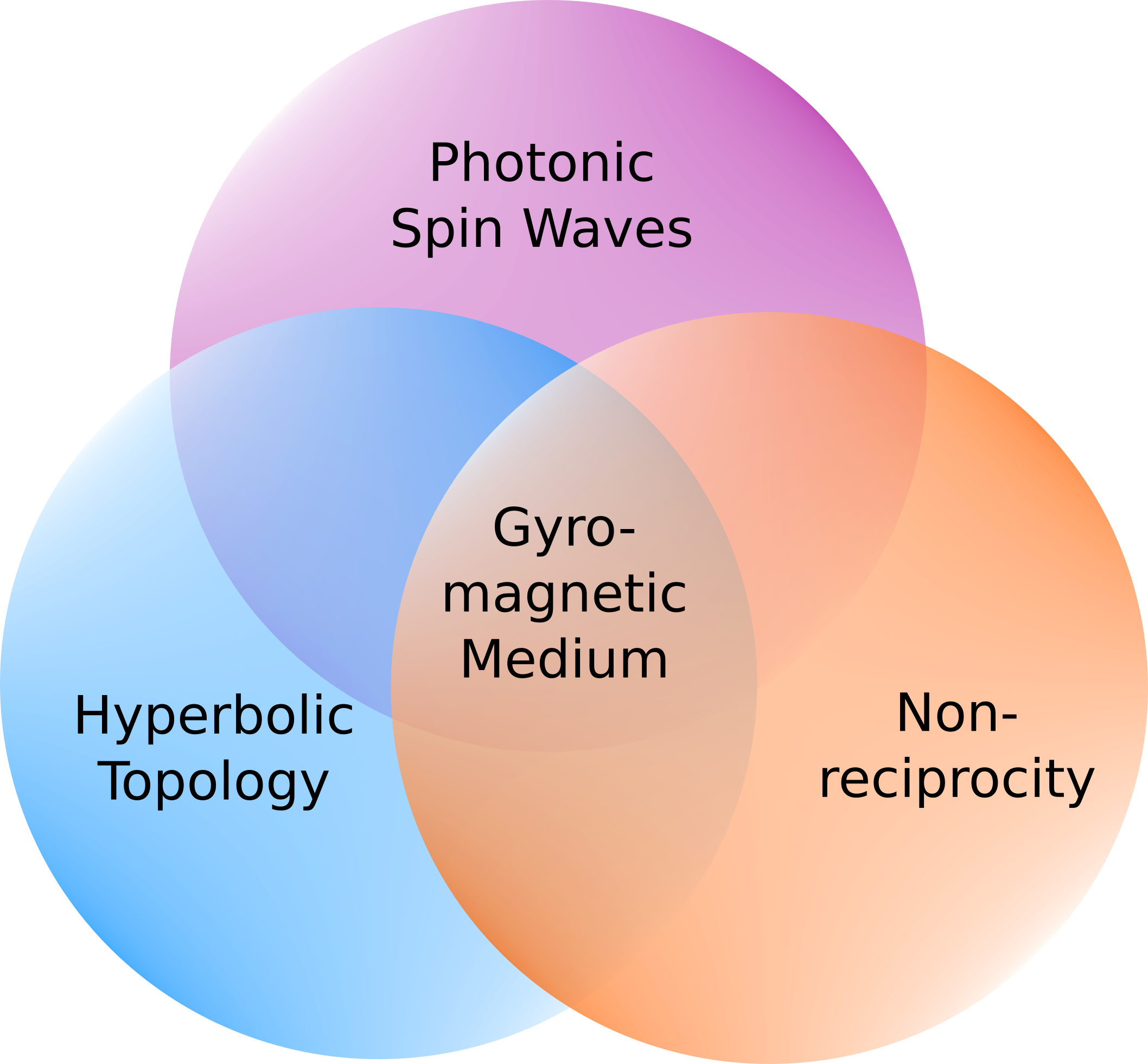}
    \caption{Gyromagnetic materials exhibit the three important, yet seemingly different phenomena of photonic spin waves, non-reciprocity, and hyperbolic topology.}
    \label{fig:fig-1}
\end{figure}

Recent progress in engineered materials with novel electromagnetic properties is paving the way for technological advances on several fronts, ranging from non-reciprocal devices \cite{non_recp_fer_2020_1,non_recp_fer_2020_2,non_recp_fer_2018_1,non_recp_fer_2019_1,non_rec_meta_surf,nnrcproc_ncom_gnt} to hyperbolic metamaterials \cite{anti_fer_mag_hyper} and photonic Chern insulators \cite{chern_insul_2020_1}. 
Three major electromagnetic wave phenomena, namely the (i) non-reciprocity, (ii) photonic spin, and (iii) hyperbolic topology, are at the forefront of next-generation devices, with all three posing unique challenges and opportunities. Interestingly, gyromagnetic materials exhibit all three phenomena (see Fig.~\ref{fig:fig-1}). Gyromagnetic materials such as ferrites have been extensively used in non-reciprocal devices at microwave frequencies \cite{dm_pozzar_book,spin_one_way_gyromag,Lokk2017_gyromag_iso_freq_surface,arnaud2020_ferrite_ant_circ_patch_LEO_sat}, with microwave isolators and circulators being their primary applications. Conventionally, ferrite-based devices were bulky and not suitable for system integration. However, recently researchers have integrated ferrite in substrate-integrated waveguides for non-reciprocal mode conversion \cite{siw_mod_conv_2019} and filter applications \cite{9238458}. The application of ferrite nanoconduits for nanometer-scale RF magnonic interconnects has also been reported recently \cite{nano_conduits_magnonics_2020}. 

Gyromagnetic materials inherently support photonic spin waves along with non-reciprocity \cite{stancil2009spin_book}. The spin of EM wave interacts with the magnetic spin of the gyromagnetic material resulting in spin waves \cite{chap_2_book_magnon}. Spin photonics is finding important applications, especially at optical frequencies \cite{spin_polariton_two_dim,magnon_photon_phonon_ent,magnon_photon_coupling}. The manifestation of photonic spin has been reported in surface polaritons \cite{phot_spin_2012_1}, spin-momentum locking in reciprocal structures \cite{todd_spin_lock,spin_mom_lock_cuprate,spin_mom_lock_nature}, and spin governed optical forces \cite{farid_spin_lock,spin_opt_forces_trans_nat,spin_opt_forces_opti_trap}.

Another important area that is at the forefront of electromagnetic materials research is hyperbolic metamaterials (HMM). Wide applications of engineered materials exhibiting hyperbolic isofrequency curves have been reported in the past \cite{hyp_meta_1,hyp_meta_2,hyp_meta_3,hyp_meta_4,hyp_meta_5,hyper_6}. Prominent applications of HMMs include super-resolution microscopy with hyperlensing \cite{hyperbolic_hyper_lens,super_lense_near_field}, enhanced near field thermal radiation \cite{therm_rad_hyp}, and threshold-less existence of Cherenkov radiation \cite{liu2017integrated,cherenkov_natur_hyper_metarial}. To achieve hyperbolic topology, conventionally, engineered anisotropic materials are fabricated, which have a negative dielectric constant along one or two directions. However, because of the complexities involved in the fabrication process, researchers have started exploring naturally occurring hyperbolic materials such as hexagonal-Boron-Nitrite \cite{hex_boron_nitride,hex_boron_nitride_nat}. 

Integration of the hyperbolic properties with the spin and non-reciprocity of gyrotropic materials is expected to result in interesting applications such as miniaturization of non-reciprocal components, unidirectional surface waves, and negative refraction, among others. Recently, researchers have reported the integration of engineered hyperbolic dielectric properties with gyromagnetic material \cite{neg_ref_fer_2015_1,anti_fer_mag_hyper,chern_insul_2020_1}. In \cite{anti_fer_mag_hyper}, the behavior of surface magnon polaritons is governed by the hyperbolic topology, which arises from a canted antiferromagnetic crystal. More recently, researchers have reported photonic Chern insulators using gyromagnetic hyperbolic metamaterial formed by a superlattice of Indium-Antimony and Yttrium Iron Garnet \cite{chern_insul_2020_1}. Gyromagnetic materials thus exhibit the three important photonic phenomena of hyperbolic topology, spin waves, and non-reciprocity as illustrated in Fig.~\ref{fig:fig-1}. However, the interplay of the three phenomena requires further investigation to enable device-level applications.

In this paper, we investigate the effect of gyrotropy on the isofrequency surfaces and spin characteristics in a gyromagnetic medium. We explore the gyrotropy-imposed conditions under which various isofrequency surfaces can exist, and reveal the nature of longitudinal and transverse spin along the 3D topological surfaces. We show that when the gyrotropy is strong, it can suppress the existence of a topological surface which would have existed in a non-gyromagnetic medium. Similarly, a strong gyrotropy can support the existence of a topological surface that would not exist in the absence of gyrotropy. We explore the spin-dependent nature of the topological surfaces, which are suppressed or supported by gyrotropy. Further, we show that the spin-profile along an isofrequency surface is locked to the direction of magnetic bias and not to the direction of propagation, resulting in violation of spin-momentum locking. We also consider a case where a conflict arises in the gyrotropy-induced spin and the spin-momentum locking, leading to an asymmetric mode profile and also a gyrotropy-induced cut-off in waveguide modes. Our work explains the spin governed nature of isofrequency surfaces and its link with the breaking of spin-momentum locking and non-reciprocity. The analysis presented in this paper will provide guiding principles for new engineered gyromagnetic materials with tailored spin and topological properties and will motivate new waveguiding structures and devices based on gyrotropy governed propagation characteristics.

% \textit{Why this work is important}.

\section{Topology in a gyromagnetic medium}
Propagation characteristics of an electromagnetic wave in a medium are governed by four-dimensional dispersion surfaces in which three dimensions correspond to the propagation vector in 3D space, and the fourth dimension corresponds to the frequency. For a fixed frequency, a dispersion surface reduces to a 3D isofrequency surface, and propagation in the medium can be characterized by the topology of the isofrequency surface. In a simple isotropic medium, the isofrequency surface has spherical topology, meaning that the propagation characteristics in the medium are uniform in all directions. In an anisotropic medium, the topology often splits into two surfaces as a consequence of different wave polarizations encountering different material properties. The two surfaces then represent the propagation characteristics for two -- often orthogonal -- wave polarizations. Engineering the topology of isofrequency surfaces is, therefore, turning out to be a powerful guiding principle for tailoring the electromagnetic properties of matter. The topological properties of non-gyrotropic medium with anisotropic permittivity tensor $\overset{\leftrightarrow}{\epsilon_r}$ have been extensively studied in the past. In recent years topological properties of magnetic materials with engineered permeability tensor are also being investigated \cite{neg_ref_fer_2015_1}, primarily because magnetic materials also exhibit gyrotropic non-reciprocity along with anisotropy. The topology of a gyromagnetic material not only depends on the permeability values but also on the gyrotropic term, and it becomes crucial to understand the effect of gyrotropy on the topology.

\begin{figure}
    \centering
    \includegraphics[width=0.8\linewidth]{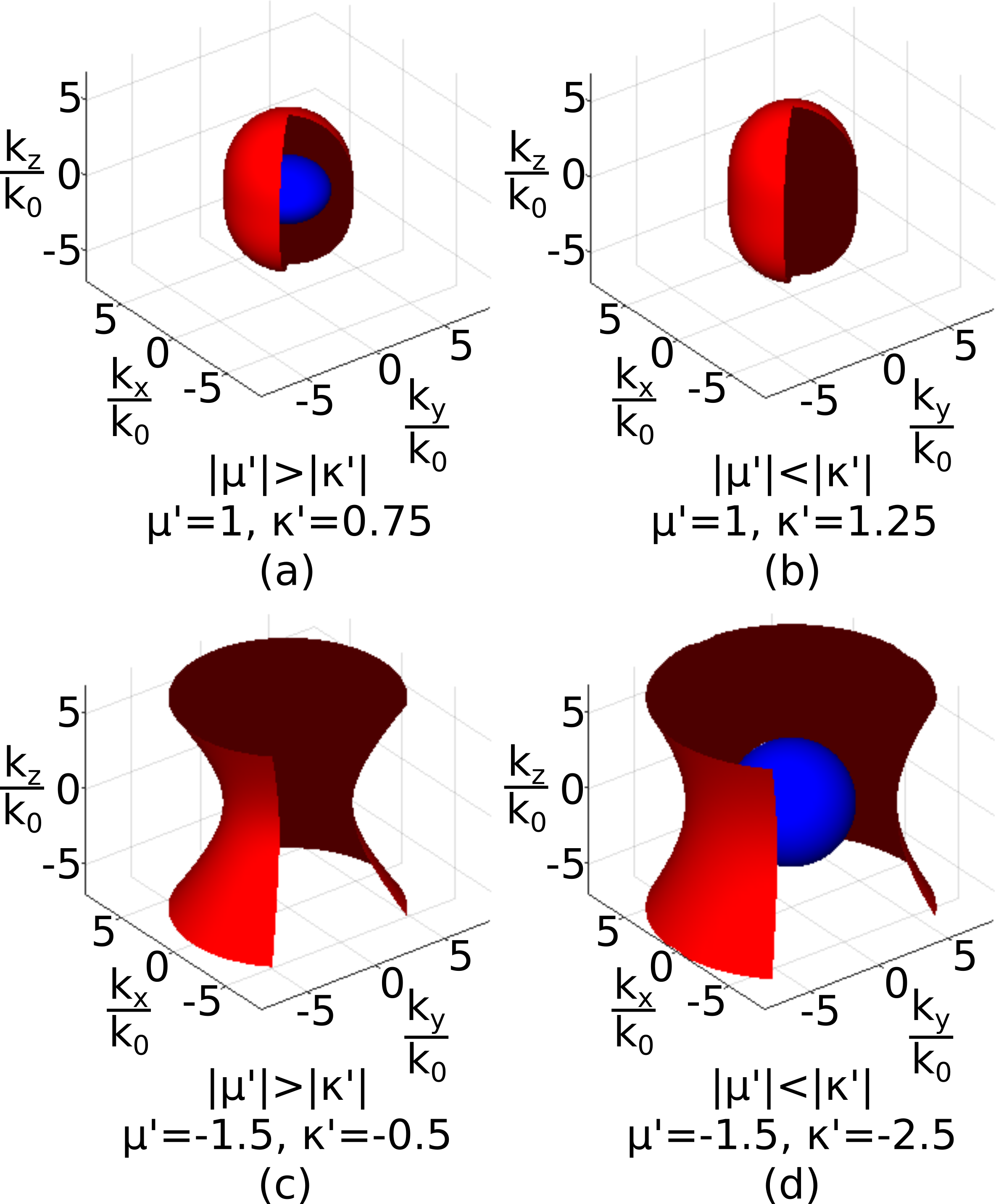}
    \caption{The topology of isofrequency surfaces in a gyromagnetic medium. Panel (a) and (b) depict the topology for $\mu^\prime=1$ when the magnitude of $\kappa^\prime$ is less than and greater then $\mu^\prime$, respectively. Panel (c) and (d) represent the topology of the isofrequency surfaces when $\mu^\prime$ is negative. (a) It can be seen that for positive $\mu^\prime$ two propagating modes exist with ellipsoidal topology when $\mu^\prime >\kappa^\prime$. (b) When $\mu^\prime <\kappa^\prime$ only one ellipsoidal surface exists. (c) For negative $\mu^\prime$, only hyperbolic mode exists when $|\mu^\prime|<|\kappa^\prime|$. (d) For negative $\mu^\prime$, an ellipsoidal topology  mode exist when $|\mu^\prime|<|\kappa^\prime|$. The dielectric constant of the medium $\epsilon_r$ is 14.}    
    \label{fig:Fig-2}
\end{figure}

\subsection{Effect of gyrotropy on topology}

Gyrotropy of a material is characterized by the presence of complex off-diagonal terms in the permittivity tensor (for gyroelectric medium) or the permeability tensor (for gyromagnetic medium). In a non-gyrotropic anisotropic medium, when all the three diagonal terms are positive (but not equal), the resulting topology consists of two concentric ellipsoids touching at specific points. When either one or two of the diagonal elements become negative, the resulting topology shifts from ellipsoidal to Type-I and Type-II hyperboloid surfaces, respectively. The topology of isofrequency surfaces is primarily governed by the diagonal terms in the material tensor. However, gyrotropy imposes additional conditions on the shape and existence of isofrequency surfaces, as will be investigated in this section.

To investigate the effect of gyrotropy on the topology, we will consider a gyromagnetic medium represented by the standard permeability tensor\cite{dm_pozzar_book}.% \cite{dm_pozzar_book},
\begin{equation}
    \label{eq:perm_tens}
    \overset{\leftrightarrow}{\mu_r}=
    \begin{bmatrix}
    \mu^\prime & -j\kappa^\prime&0\\
    j\kappa^\prime&\mu^\prime&0\\
    0&0&1\\
    \end{bmatrix}.
\end{equation}
Here $\mu^\prime$ is the relative permeability along $\hat{x}$ and $\hat{y}$ directions. $\kappa^\prime$ is the gyrotropic term. Relative permeability along the $\hat{z}$ direction is 1. This anisotropic permeability tensor $\overset{\leftrightarrow}{\mu_r}$ in eq. (\ref{eq:perm_tens}) corresponds to a ferrite material with DC magnetic bias along $+\hat{z}$ direction. We assume the material to be electrically isotropic with relative permittivity tensor $\overset{\leftrightarrow}{\epsilon_r}=\epsilon_r[I_3]$, where $[I_3]$ is a $3\times3$ identity matrix, and $\epsilon_r$ is the relative dielectric constant. 

Wave propagation in this gyromagnetic medium is governed by,
\begin{equation}
    \label{eq:elm_h}
    [\overset{\leftrightarrow}{k}\cdot\overset{\leftrightarrow}{k}+k_0^2\epsilon_r\overset{\leftrightarrow}{\mu_r}]\cdot\Vec{H}=0
\end{equation}
where $k_0=\omega\sqrt{\epsilon_0\mu_0}$ is the free space wave vector, $\omega$ is the angular frequency, $\epsilon_0$ and $\mu_0$ are the absolute permittivity and permeability of free space, respectively. $\overset{\leftrightarrow}{k}$ represents curl operation in the matrix form given by,
\begin{equation}
    \label{eq:k_ten}
    \overset{\leftrightarrow}{k}=\begin{bmatrix}
    0&-k_z&k_y\\
    k_z&0&-k_x\\
    -k_y&k_x&0\\
    \end{bmatrix}.
\end{equation}
$k_x$, $k_y$, and $k_z$ in the above equation represent the propagation constant along $\hat{x}$, $\hat{y}$, and $\hat{z}$ directions, respectively. By equating the determinant of the matrix $\left[\overset{\leftrightarrow}{k}\cdot\overset{\leftrightarrow}{k}+k_0^2\epsilon_r\overset{\leftrightarrow}{\mu_r}\right]$ to zero, for a fixed frequency $\omega$, we get the equation of isofrequency surface, as  
\begin{multline}
    \label{eq:num_sol_eq}
    (k_x^2+k_y^2+k_z^2)(k_z^2+(k_x^2+k_y^2)\mu^\prime)+\epsilon_r^2k_0^4(\mu^{\prime2}-\kappa^{\prime2})\\+\epsilon_rk_0^2(\kappa^{\prime2}(k_x^2+k_y^2)-\mu^\prime(2k_z^2+(k_x^2+k_y^2)(1+\mu^\prime)))=0.
\end{multline}  %%%%%%
The topology of the isofrequency surface can be computed by solving Eq.~(\ref{eq:num_sol_eq}) in the three-dimensional k-space. Note that Eq.~(\ref{eq:num_sol_eq}) is bi-quadratic in $k_x$, $k_y$, and $k_z$, indicating a possibility of the existence of two topological surfaces. The two topological surfaces correspond to two different wave spin or polarizations. However, depending on the values of $\mu^\prime$ and $\kappa^\prime$, only one of the surface may exist in real-valued k-space. To understand the effect of gyrotropic term on the topology, we take four cases for the components $\mu^\prime$ and $\kappa^\prime$ to obtain different topological regimes of isofrequency surfaces as shown in Fig.~\ref{fig:Fig-2}.   

For a positive value of $\mu^\prime$, in a non-gyromagnetic ($\kappa^\prime=0$) medium, we would expect the existence of two concentric ellipsoidal surfaces, touching one another at specific points. The contact points would indicate that the two wave spins represented by the two isofrequency surfaces have momentum degeneracy in certain directions. However, in the presence of gyrotropy ($\kappa^\prime \neq 0 $) with the magnitude of the gyrotropic term less than the permeability term, i.e., $|\mu^\prime|>|\kappa^\prime|$, the topology consists of two completely non-contact concentric ellipsoids as observed in Fig.~\ref{fig:Fig-2}(a). As we increase the magnitude of $\kappa^\prime$, the effects of gyrotropy start playing a more prominent role, and when $|\kappa^\prime|>|\mu^\prime|$ only one ellipsoidal isofrequency surface exists, as plotted in Fig.~\ref{fig:Fig-2}(b). Gyrotropy, in this case, suppresses the existence of one type of wave spin and limits the topology of the medium to a single surface.  

Similarly, when $\mu^\prime$ is negative in a non-gyromagnetic medium ($\kappa^\prime=0$), we would expect a single Type-II hyperboloid. The negative diagonal terms in the permeability tensor would suppress the existence of the second surface. In the presence of gyrotropy, as long as the gyrotropic term is less than the permeability term, i.e., $|\mu^\prime|>|\kappa^\prime|$, there exists a single hyperbolic isofrequency surface as in Fig.~\ref{fig:Fig-2}(c). However,  when the magnitude of gyrotropic term exceeds the magnitude of permeability term, i.e., $|\mu^\prime|<|\kappa^\prime|$, we get two isofrequency surfaces, one hyperboloid  and the second ellipsoid, shown in Fig.~\ref{fig:Fig-2}(d). In this case, the gyrotropy supports the existence of the second ellipsoidal surface, which otherwise would be non-existent. These conditions are summarized in Table \ref{tab:up_kp_vals}.

\begin{table}[]
    \centering
    \caption{Topological regimes for different values of $\mu'$ and $\kappa'$, as depicted in Fig. \ref{fig:Fig-2}.}
    
    \begin{tabular}{|m{0.2\linewidth}|m{0.4\linewidth}|m{0.4\linewidth}|}
        \hline
        \textbf{Gyrotropic term} & $\mu^\prime>0$ & $\mu^\prime<0$  \\
        \hline 
        \rowcolor{LightCyan}
        $\kappa^\prime=0$ & Two concentric touching ellipsoids & Single hyperboloid \\ \hline
        \hspace{1.9mm}$|\mu^\prime|>|\kappa^\prime|$ & Two concentric non-touching ellipsoids & Single hyperboloid \\
        \hline
        \hspace{1.9mm}$|\mu^\prime|<|\kappa^\prime|$ & A single ellipsoid & One hyperboloid and one ellipsoidal \\
        \hline
        \rowcolor{LightCyan}
    \end{tabular}
    \label{tab:up_kp_vals}
\end{table}

While the nature (ellipsoidal or hyperboloid) of the topology depends on the permeability terms, gyrotropy also imposes a strong influence on the topology. This influence is exerted by the interaction of gyromagnetic material spin with the photonic spin of the wave and forms the subject of investigation in Section~\ref{section-spin-waves}. Before we delve deeper into the nature of spin waves along the topological surfaces, let us first look at the topological behavior in natural gyromagnetic material and identify its different topological regimes.

%%%%%%%%%%%%%%%%%

\subsection{Topological regimes in natural gyromagnetic medium}

\begin{figure}
    \centering
    \includegraphics{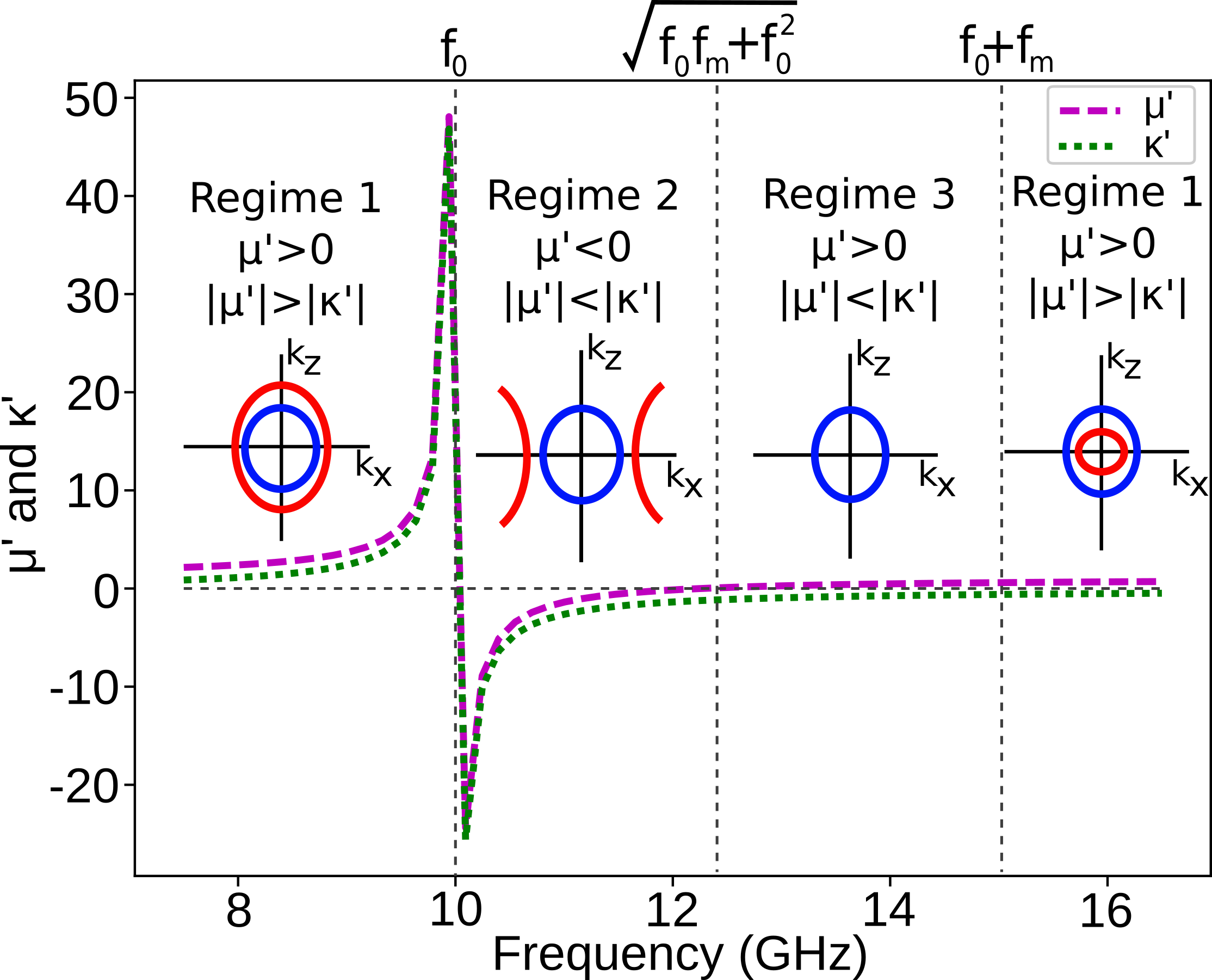}
    \caption{Variation of components of permeability tensor $\overset{\leftrightarrow}{\mu_r}$ in YIG, with change of frequency. Regimes are classified as per the topology of isofrequency contours, which depends on the values of $\mu'$ and $\kappa'$. For regime-1, both the isofrequency surfaces are ellipsoidal, both $\mu^\prime$ and $\kappa^\prime$ are positive. Regime-1 extends up to the Larmor resonance point, beyond which mode 2 isofrequency contour becomes hyperbolic, and mode 1 isofrequency contour remains ellipsoidal, both $\mu^\prime$ and $\kappa^\prime$ are negative. Regime-3 starts at $\sqrt{f_0f_m+f_0^2}$ and ends at $f_0+f_m$. Only mode 1 isofrequency surface exists in this region, $\mu^\prime>0$, $\kappa^\prime<0$ and $|\mu^\prime|<|\kappa^\prime|$ . Beyond regime-3, conditions for $\mu'$ and $\kappa'$ match to that of regime-1, and both the isofrequency surfaces are ellipsoidal in nature.}
    \label{fig:up_kp_vs_freq}
\end{figure}

\begin{figure}
    \centering
    \includegraphics[scale=1]{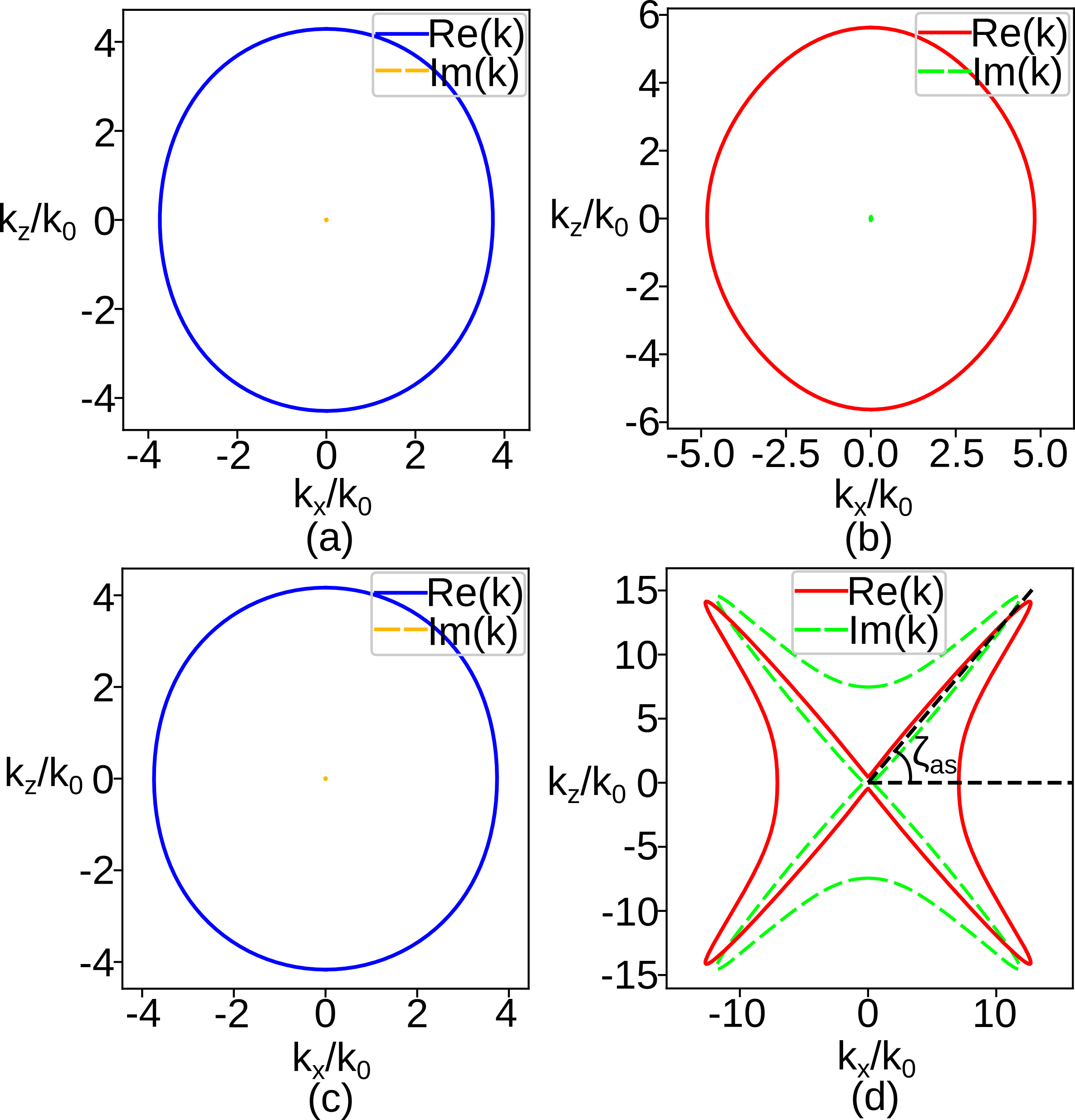}
    \caption{Real and imaginary component of isofrequency contours representing propagating and attenuating solutions, respectively, in YIG with material losses. Panel (a) and (b) shows real and imaginary values of the isofrequency surface at 6 GHz,and correspondingly panel (c) and (d) shows for 11 GHz, respectively. $\Delta_H=75 Oe$ and $\epsilon^{\prime\prime}=2\times10^{-4}$. $\upzeta_{as}$ is the angle of asymptote for the hyperbola. Singularity is restricted along the asymptote due to the inclusion of losses. The overall effect of loss is insignificant in the topology.}
    \label{fig:pol_6_11_cnt}
\end{figure}

Elements of the permeability tensor in a natural gyromagnetic medium such as ferrite are given by $\mu^\prime=1+f_0f_m/(f_0^2-f^2)$ and $\kappa^\prime=ff_m/(f_0^2-f^2)$.
Here $f_0$ is Larmor precession frequency and is directly associated with the applied DC magnetic bias \cite{dm_pozzar_book}. Frequency parameter $f_m$ depends on the saturation magnetization of the ferrite. For our study, we consider Yttrium Iron Garnet (YIG), a ferrite of the Garnet class, which is one of the most commonly used ferrite material at microwave frequencies due to its significantly low loss and very narrow ferromagnetic resonance linewidth\cite{CHENG20183018}. We use bias field strength $H_0$ of 3570 Oe along $\hat{z}$ and saturation magnetization $4\pi M_s$ as 1800 Gauss, which gives $f_0=9.99 GHz$ and $f_m=5.04 GHz$. 
$\mu^\prime$ and $\kappa^\prime$ depend on the bias strength and saturation magnetization and also on the frequency of operation.

Figure~\ref{fig:up_kp_vs_freq} shows the values of $\mu^\prime$ and $\kappa^\prime$ as a function of frequency. Variation in the values of $\mu^\prime$ and $\kappa^\prime$ gives rise to three different topological regimes. Regime-1 and regime-2 are separated at the Larmor resonance frequency $f_0$. At the Larmor frequency, $\mu^\prime$ and $\kappa^\prime$ undergo singularity, and there is a sudden change in their sign from positive to negative. In regime-1, $\mu^\prime>0$, $\kappa^\prime>0$ with $|\mu^\prime|>|\kappa^\prime|$. This results in the existence of two  modes with elliptical topology. In regime-2, $\mu^\prime<0$, $\kappa^\prime<0$ and $|\mu^\prime|<|\kappa^\prime|$. Therefore as per our discussion in the previous subsection, one of the modes in this region attains hyperbolic topology while the other is elliptical. This elliptical mode would not have existed if the magnitude of $\kappa^\prime$ was smaller than the magnitude of $\mu^\prime$.

The frequency at which $\mu^\prime$ becomes positive from negative marks the end of regime-2 and the start of regime-3. This crossover frequency can be calculated as $\sqrt{f_0f_m+f_0^2}$. In this regime-3, the condition $|\mu^\prime|<|\kappa^\prime|$ is maintained, which results in only a single elliptical (since $\mu^\prime>0$) surface, with the other surface being suppressd by gyrotropy. Regime-1 reappears at the frequencies $>f_0+f_m$ where $\kappa^\prime$ is still negative but $|\mu^\prime|>|\kappa^\prime|$. Beyond regime-3, in regime-1, the mode suppressd by gyrotropy reemerges as an elliptical isofrequency contour.

A natural gyrotropic medium is also expected to have some material losses. To ensure that the losses do not significantly effect the topology and the subsequent discussions in this paper, we also compute the topology in YIG material while including the material losses. Figure~\ref{fig:pol_6_11_cnt} shows the two modes for frequency 6 GHz and 11 GHz representing regime-1 and regime-2, respectively.
Two significant agents of losses are dielectric losses and gyromagnetic losses. Dielectric losses and gyromagnetic losses are included in the expressions by using complex values of relative permittivity as $\epsilon_r=\epsilon'+j\epsilon''$, where $\epsilon'=14$ and $\epsilon''=2\times10^{-4}$ and Larmor frequency $f_0$ as $f_0=f_0-j\alpha$, where $\alpha=\mu_0\gamma\Delta_H\times10^3/16\pi^2$ \cite{dm_pozzar_book}.
Linewidth, $\Delta_H $, of the YIG is $\Delta_H=75 Oe$. 
Figure~\ref{fig:pol_6_11_cnt}(a) and (b) shows real and imaginary values of the two modes of the isofrequency contour at 6 GHz, and Fig.~\ref{fig:pol_6_11_cnt}(c) and (d) shows real and imaginary values for the modes at 11 GHz.
Except for the hyperbolic case, the imaginary values are always relatively low in magnitude, appearing as a dot at the center for the elliptical curve. For the hyperbolic case, losses restrict the hyperbolic singularity along the asymptote angle. Moreover, for the hyperbolic mode at 11 GHz, the real and imaginary values are comparable only at close proximity to the angle of asymptote $\upzeta_{as}$. This observation allows us to ignore the losses of the medium for further analysis, as their contribution to the overall topology is relatively insignificant. 

In the next section, we investigate the nature of the spin-profile of the topological surfaces.

%%%%%%%%%%%%%%%%%%

\section{Spin-profile of topological surfaces in gyromagnetic medium}\label{section-spin-waves}

\begin{figure}
    \centering
    \includegraphics{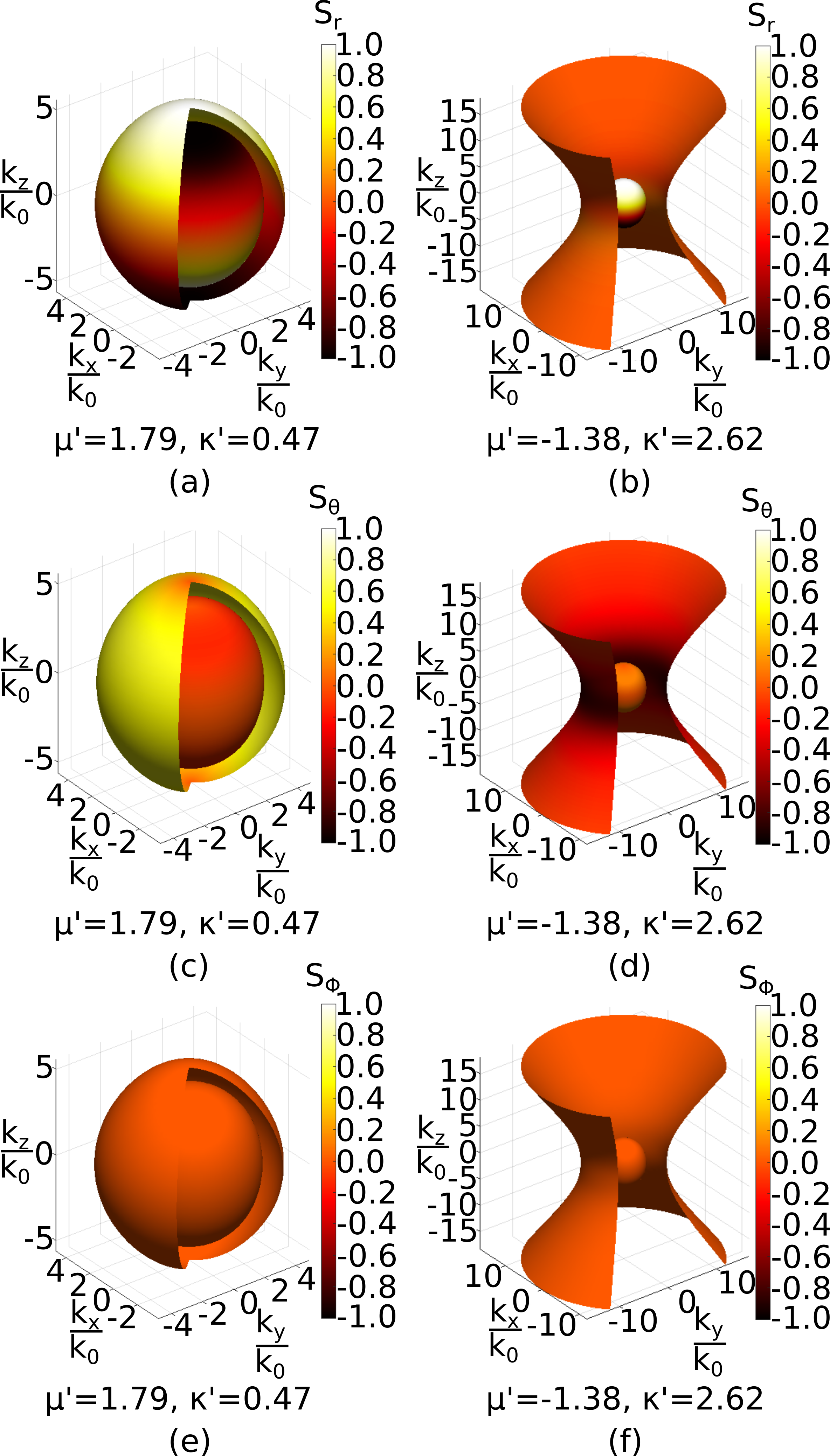}
    \caption{Representation of spin-profile on the isofrequency surface for both elliptical and hyperbolic modes representing regime-1 and regime-2. Panel (a) and (b) shows the spin-profile along $\hat{r}$ plane, (c) and (d) shows spin-profile representation for $\hat{\theta}$ and (e) and (f) shows spin-profile for $\hat{\phi}$ plane, respectively.}
    \label{fig:3d_hot_surface}
\end{figure}

Gyrotropy in a magnetic medium is introduced due to the gyration of magnetic dipoles under the influence of a DC magnetic bias. For a $z$-biased medium, the dipoles gyrate in the $x-y$ plane with material magnetic spin aligned with the magnetic bias. The gyrotropic term, $\kappa^\prime$,  therefore couples the $x$ and $y$ directional magnetic field, which results in an introduction of spin in the electromagnetic waves interacting with it \cite{stancil2009spin_book}. We will call this spin the material-induced spin. In this section, we investigate the spin-profile of the isofrequency surfaces and reveal the spin-dependent nature of gyrotropy-imposed conditions on the topology.

Since we are considering a gyromagnetic medium, we are concerned with the magnetic spin defined by $\Vec{S} = \Im\{\vec{H}^*\times\vec{H}\}$. This definition of spin is equivalent to defining the Third Stokes parameter in the three directions, with $\vec{S} = S_x\hat{x}+ S_y\hat{y}+ S_z\hat{z}$ (see Appendix B). Stokes parameters are commonly used to define polarization at the microwave frequencies. To compute the magnetic spin, the magnetic vector components $H_y$ and $H_z$ are expressed in terms of $H_x$, using the condition $[\overset{\leftrightarrow}{k}\cdot\overset{\leftrightarrow}{k}+k_0^2\epsilon_r\overset{\leftrightarrow}{\mu_r}]\cdot\Vec{H}=0$. The values of $k_x$, $k_y$, and $k_z$ are constrained by the topological surfaces obtained by $\det[\overset{\leftrightarrow}{k}\cdot\overset{\leftrightarrow}{k}+k_0^2\epsilon_r\overset{\leftrightarrow}{\mu_r}]=0$.  To plot the spin-profile of the topological surfaces, the fields are converted from Cartesian $(H_x,H_y,H_z)$ to polar $(H_r,H_\theta, H_\phi)$ using the transformation matrix,

\begin{equation}
    \begin{split}
        \begin{bmatrix}
        H_r\\
        H_\theta\\
        H_\phi\\
        \end{bmatrix}
        =
        \begin{bmatrix}
        \sin\theta\cos\phi&\sin\theta\sin\phi&\cos\theta\\
        \cos\theta\cos\phi&\cos\theta\sin\phi&-\sin\theta\\
        -\sin\phi&\cos\phi&0\\
        \end{bmatrix}
        \cdot
        \begin{bmatrix}
        H_x\\H_y\\H_z\\
        \end{bmatrix}
    \end{split}
\end{equation}
and the Cartesian wave vector $(k_x,k_y,k_z )$ is converted to a spherical coordinate representation of $(k_r,\theta,\phi)$. The three components of the spin are then computed in the spherical coordinates as $\vec{S} = S_r\hat{r} + S_\theta\hat{\theta} + S_\phi\hat{\phi}$. A wave traveling in an arbitrary direction in a gyromagnetic medium can, in general, have all three field components, and therefore we will require three components to represent the spin. $S_r$ is the longitudinal spin, while $S_\theta$ and $S_\phi$ are the transverse spins. Values of spin components may extend from -1 to 1, where the extreme values represent orthogonal circularly polarized spins while the value zero corresponds to the linear polarisation of the field with zero spin.

Figure~\ref{fig:3d_hot_surface} shows the mapping of spin $\vec{S}$ along the topological surfaces in two topological regimes. Figure~\ref{fig:3d_hot_surface}(a) shows the radial spin $S_{r}$ when $\mu^\prime>0$ and $|\mu^\prime|>|\kappa^\prime|$, while (b) shows $S_r$ when $\mu^\prime<0$ and $|\mu^\prime|<|\kappa^\prime|$. These two cases correspond to the Regime-1 and Regime-2 of the gyromagnetic medium discussed in the previous section, with the only difference that here $\kappa^\prime$ is positive. Similarly, Fig.~\ref{fig:3d_hot_surface}(c) and (d) represent $S_{\theta}$, and Fig.~\ref{fig:3d_hot_surface}(e) and (f) show the spin $S_{\phi}$ plane in the two topological regimes, respectively. From Fig.~\ref{fig:3d_hot_surface}(a), $S_r$ evidently reaches its extreme values along the positive and negative $z$-axis (the axis of bias), with the two topological surfaces supporting opposite spins. Note from Fig.~\ref{fig:3d_hot_surface}(b) that one surface is hyperbolic with $S_r$ zero, indicating the absence of longitudinal spin on the hyperbolic surface. It is also observed  from Fig.~\ref{fig:3d_hot_surface}(e,f) that no spin exists for the $S_\phi$ for both regimes. This is because the magnetic field component of $r$ and $\theta$ comes out to be in-phase and the spin in the $\hat\phi$ direction is always zero, i.e., $S_\phi=0$.

\begin{figure}
    \centering
    \includegraphics[scale=0.8]{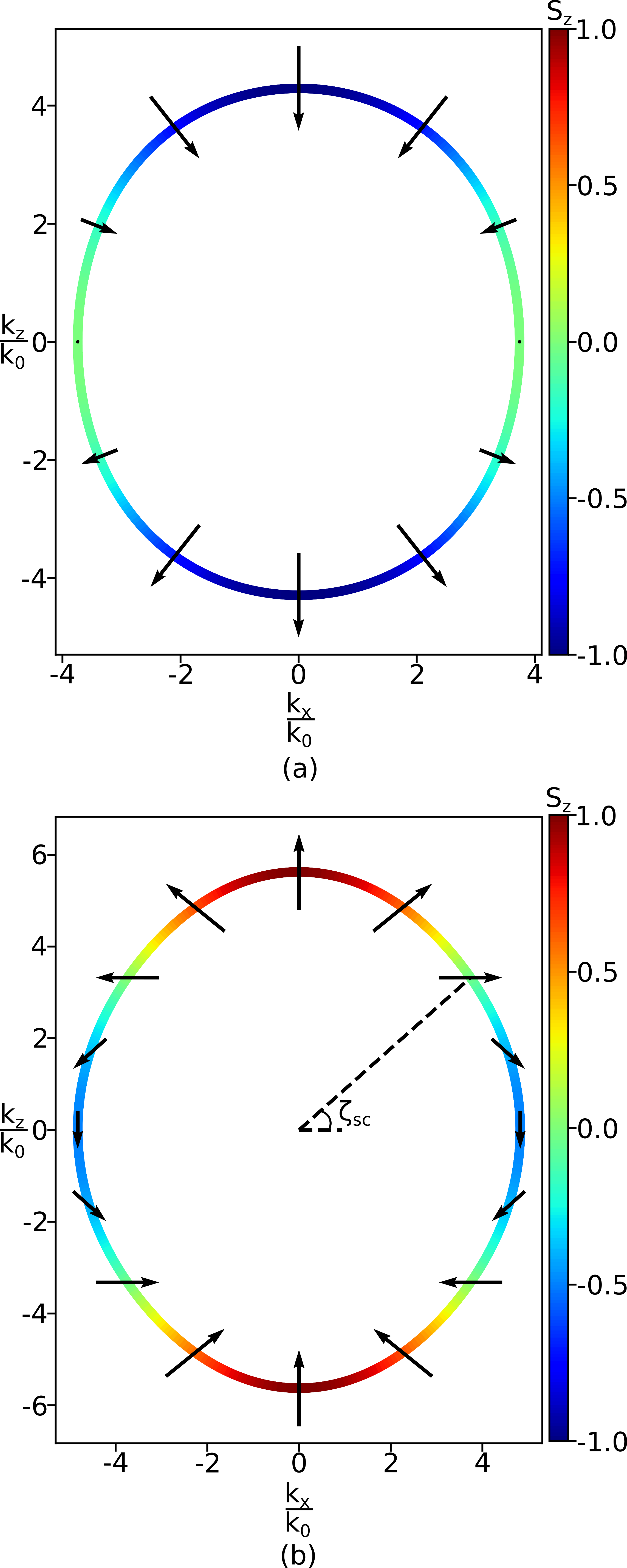}
    \caption{Representation of spin orientation along the isofrequency contour for $\mu^\prime=1.79$ and $\kappa^\prime=0.47$. The color of the isofrequency contour represents the magnitude of $\hat{z}$ spin, which is the direction of bias. The arrows signify the overall spin magnitude in the $x-z$ plane. Panel (a) and (b) depict the unique spin-profile of both the existing isofrequency contours. The spin of the mode in (a) is anti-parallel to the direction of material spin. This mode ceases to exist when $\kappa^\prime>\mu^\prime$}
    \label{fig:color_coded_spin_ncm}
\end{figure}

\begin{figure}
    \centering
    \includegraphics{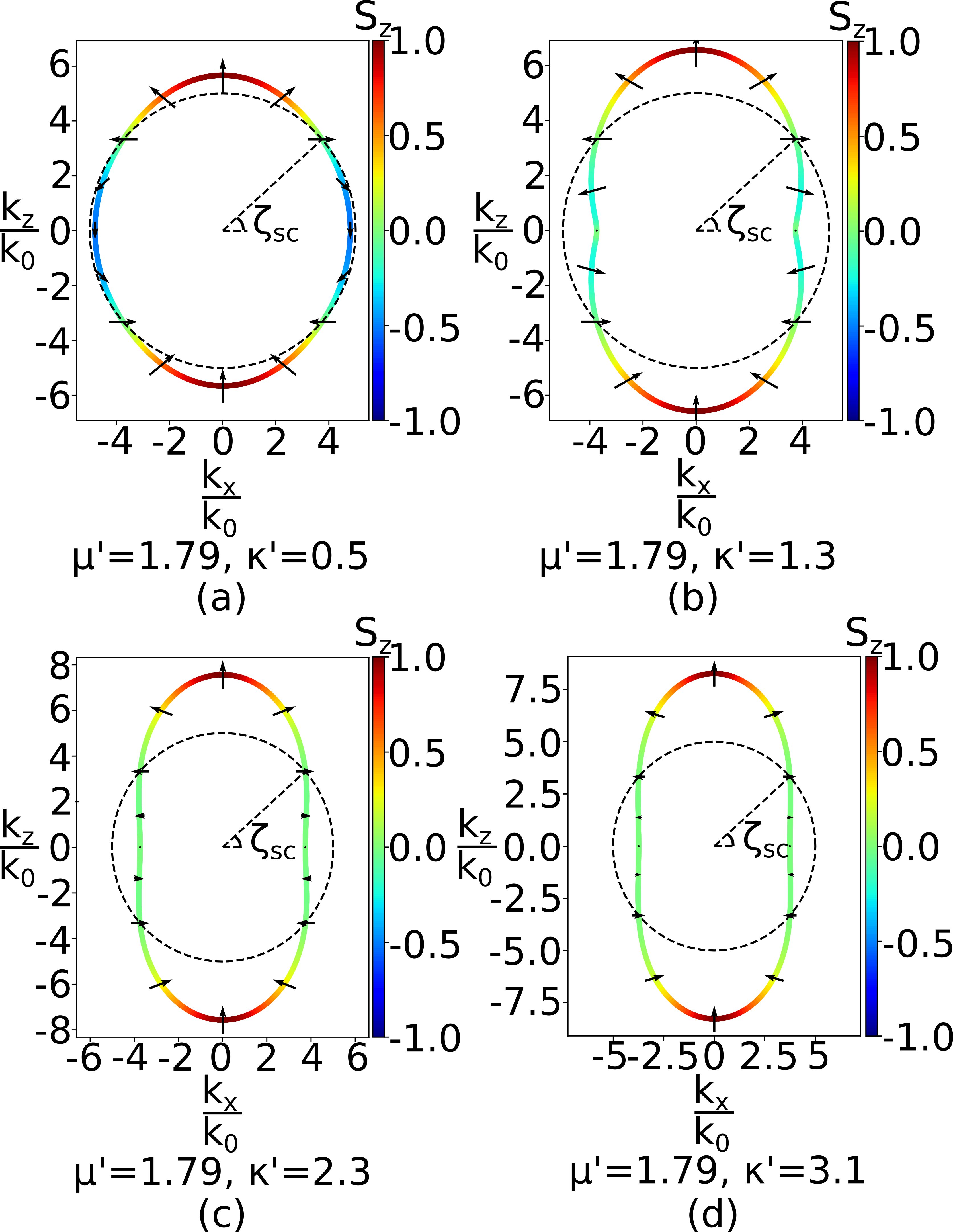}
    \caption{Variation of spin-profile with increase in gyrotropic term $\kappa^\prime$. The component of spin along the direction of bias is zero at spin-crossover angle $\zeta_{sc}$. This point is given by the intersection of isofrequency contour with gyrotropy and $k_x^2+k_z^2 = k_0^2\sqrt{\mu^\prime\epsilon_r}$.  We observe that the spin-profile is parallel to the material spin when the points are outside the circle and vice versa, for $\kappa^\prime<\mu^\prime$. When $\kappa^\prime>\mu^\prime$ the anti-parallel spin approaches zero even inside the circle.}
    \label{fig:spin_var}
\end{figure}

\begin{figure}
    \centering
    \includegraphics[scale=0.8]{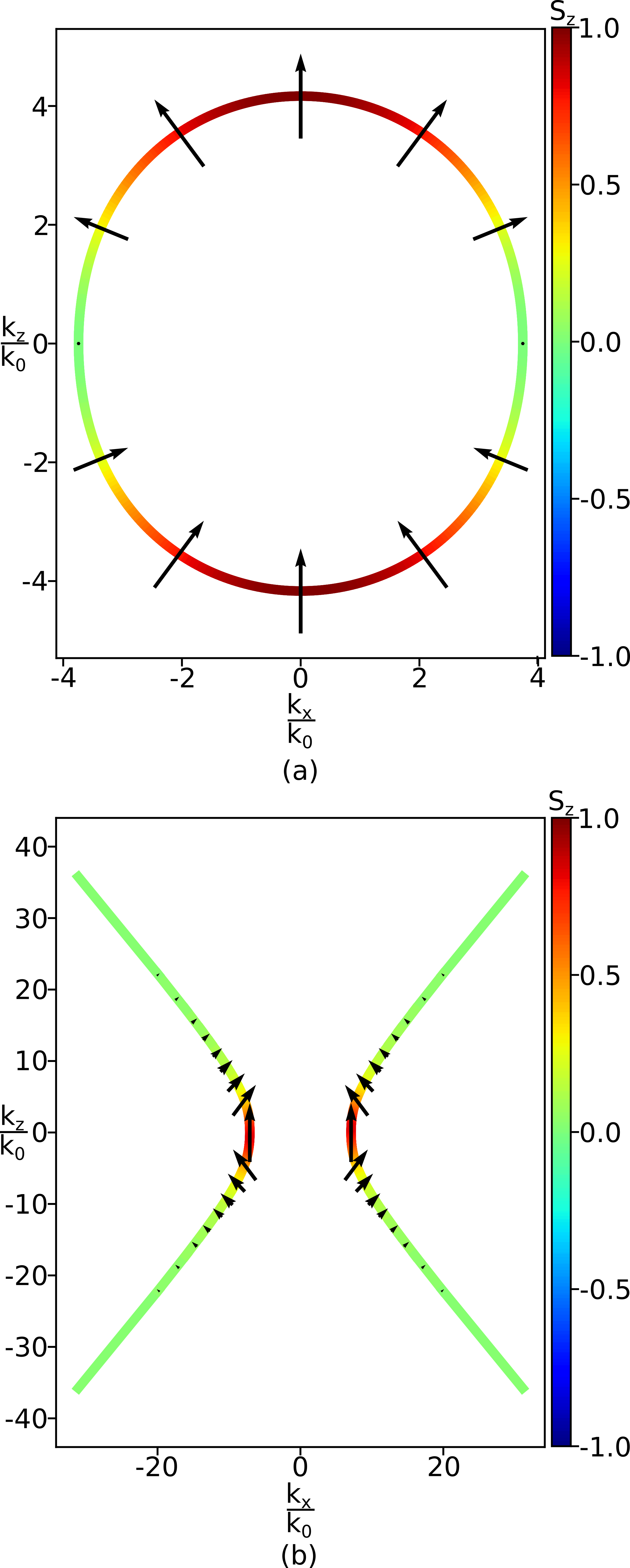}
    \caption{Representation of spin orientation along the isofrequency contour for $\mu^\prime=-1.38$ and $\kappa^\prime=2.62$. The color of the isofrequency contour represents the magnitude of $\hat{z}$ spin. The arrows signify the overall spin magnitude through  third stokes parameter $S_3$. Panel (a) shows the spin orientation for the elliptical mode, (b) Spin orientation representation for hyperbolic mode. It is observed that $\hat{z}$ spin is maximum along $k_x$ axis and negligible near the asymptote.}
    \label{fig:color_coded_spin_cm}
\end{figure}

To further investigate the nature of spin and gyrotropy imposed conditions in different topological regimes, we compute the spin along the isofrequency surfaces while restricting the wave propagation to the $x-z$ plane. We can analyze the spin-profile along the in-plane isofrequency curves without loss of generality because there is symmetry in the $x-y$ plane. For in-plane propagation, the $y$ and $z$ components of the magnetic field can be written in terms of $H_x$ in the form of analytical expressions, given by, 
\begin{equation}
    \label{eq:hy_rel}
    H_y=-\frac{j\epsilon_rk_0^2\kappa^\prime}{\epsilon_rk_0^2\mu^\prime-k_r^2}H_x,
\end{equation}
and
\begin{equation}
    \label{eq:hz_rel}
    H_z=-\frac{k_r^2\cos\theta\sin\theta}{\epsilon_rk_0^2-k_r^2\sin^2\theta}H_x.
\end{equation}
From Eq. (\ref{eq:hy_rel}) and Eq. (\ref{eq:hz_rel}), it can be seen that $x$ and $z$ components of the magnetic field are in phase while the $y$ component is $90^\circ$ out-of-phase. Note that $\kappa^\prime$ appears in the numerator of Eq (\ref{eq:hy_rel}) as a linear term, indicating the presence of a material-induced spin in the EM wave which reverses its direction with the sign of $\kappa^\prime$. 

Spins for the two modes in topological Regime-1 are computed and shown along the isofrequency curves in Fig.~\ref{fig:color_coded_spin_ncm}. The arrows represent the direction of net spin, and the colormap of the contour represents the spin $S_z$, i.e., spin in the direction of the magnetic bias. It can be seen that for the mode shown in  Fig.~\ref{fig:color_coded_spin_ncm}(a), the spin along $\hat{z}$ direction is anti-parallel to the material spin ($+\hat{z}$). Here we would like to highlight that this mode with spin component anti-parallel to material spin would not exist if $|\kappa^\prime|>\mu^\prime$, i.e. when the gyrotropic term is strong enough to suppress the topological surface which has an anti-parallel spin component. Thus spin plays a crucial role in deciding the existence of the topological surface shown in Fig.~\ref{fig:color_coded_spin_ncm}(a).

The mode represented in Fig.~\ref{fig:color_coded_spin_ncm}(b) has a spin component parallel to the direction of material spin when the angle of propagation from the $k_x$-axis is greater than $\upzeta_{sc}$, while the wave spin has a component anti-parallel to the direction of material spin when the angle of propagation (w.r.t. $x$-axis) is less than $\upzeta_{sc}$. We call $\upzeta_{sc}$ as the spin-crossover angle. At $\upzeta_{sc}$, the wave has zero spin along the direction of the magnetic bias. Interestingly the spin-crossover angle $\upzeta_{sc}$ is independent of the gyrotropic term $\kappa^\prime$ and is always given by (see Appendix C),
\begin{equation}
    \label{eq:theta_s}
    \upzeta_s=\cos^{-1}\sqrt{\frac{1}{\mu^\prime}}.
\end{equation}
The spin-crossover point can be physically interpreted as the intersection of the isofrequency curve in presence ($\kappa^\prime\neq0$) of gyrotropic term with the curve $k_x^2+k_z^2 = k_0^2\sqrt{\mu^\prime\epsilon_r}$. Figure~\ref{fig:spin_var} shows the spin-profile along the isofrequency surface for increasing values of $\kappa^\prime$. It can be seen that for all the cases, the spin-crossover point is given by the intersection of the gyromagnetic isofrequency curve and the non-gyromagnetic curve $k_x^2+k_z^2 = k_0^2\sqrt{\mu^\prime\epsilon_r}$ (shown by the dashed curves). An increase in the magnitude of the gyrotropic term ($\kappa^\prime$) results in an elongation of the topological surface along the $k_z$ axis. It can be further noted that as the value of $\kappa^\prime$ increases, the magnitude of spin anti-parallel to the magnetic bias decreases and approaches zero even for angles less than $\upzeta_{sc}$.

The role of spin in the topological properties can be further corroborated by investigating the spin-profile in the Regime-2 with $\mu^\prime<0$ and $|\mu^\prime|<|\kappa^\prime|$, which supports a hyperbolic and an elliptical mode. The spin-profile of the two modes in Regime-2 is shown in Fig.~\ref{fig:color_coded_spin_cm}. The elliptical mode has a spin component parallel to the direction of material spin. This mode would not have existed if the magnitude of $\kappa^\prime$ was less than the magnitude of $\mu^\prime$. Here the gyrotropy is strong enough to support a spin mode which the negative terms in anisotropic permeability would have suppressed. The hyperbolic mode also shows interesting spin characteristics. The hyperbolic mode has zero longitudinal (radial) spin throughout the curve, and the net spin also goes to zero along the asymptote. Near the $k_x$ axis, the spin is aligned parallel to the magnetic bias.  

In all cases discussed above, the spin for a pair of antipodal points on the topological surface is aligned in the same direction.  This indicates that the spin for two counter-propagating momentum vectors $\vec{k}$ is aligned in the same direction, which is in contrast to a non-gyromagnetic medium where the spin is reversed with the direction of propagation. Here the spin is locked to the material instead of the momentum. The non-reversal of spin with the direction of propagation is a result of the breaking of time-reversal symmetry (TRS) in a gyromagnetic medium. In the next section, we discuss the effect of broken TRS on spin-momentum locking.

\section{Breaking of time-reversal symmetry and spin-momentum locking}
\begin{figure}
    \centering
    \includegraphics[width=\linewidth]{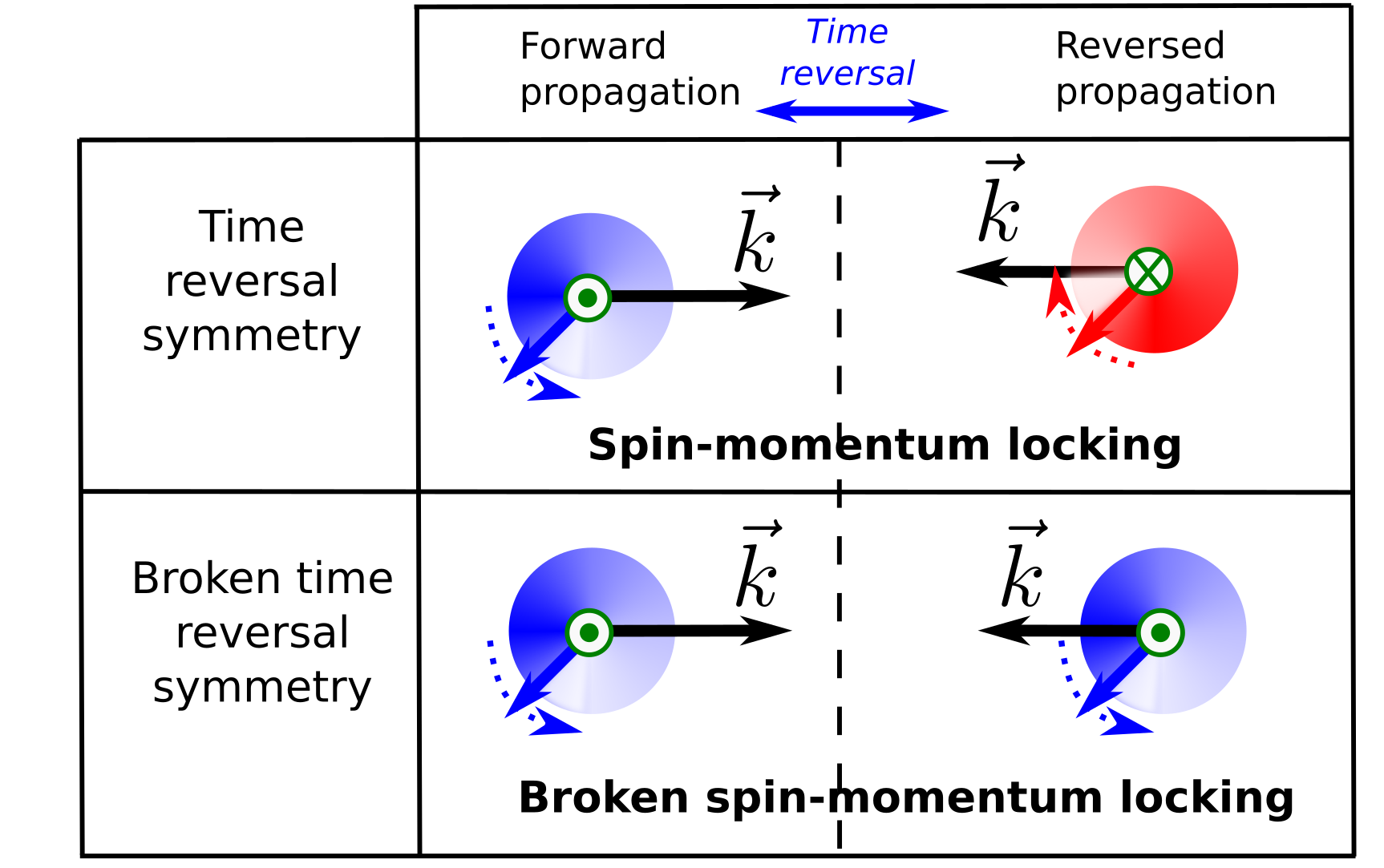}
    \caption{In a medium with time-reversal symmetry, the direction of momentum as well as the rotation of fields is reversed with the reversal of time. Therefore spin-momentum symmetry emerges as a consequence of TRS. However, in a gyromagnetic medium, the spin is locked to the medium resulting in the breaking of TRS and spin-momentum locking.}
    \label{fig:fig6_sec4}
\end{figure}

Spin emerges as a consequence of the rotation of electric or magnetic field vectors with time. In a system with TRS, as time is reversed, the direction of both spin, as well as momentum, gets reversed, resulting in a spin-momentum locking (see Fig.~\ref{fig:fig6_sec4}). However, in non-reciprocal systems where the TRS is broken, the spin-momentum locking may not be maintained. In \cite{Pendharker:18}, Pendharker et al. showed that the spins of two counter-propagating waves are opposite but not equal (non-degenerate) as the TRS is broken by motion-induced non-reciprocity. A gyromagnetic medium, on the other hand, exhibits material-induced non-reciprocity where the two counter-propagating waves have equal and aligned spins, as shown in the previous section. In a gyromagnetic medium, the spin is locked to the material, such that the reversal of time does not reverse the rotation of the magnetic field while the direction of propagation is reversed, as depicted in Fig.~\ref{fig:fig6_sec4}. This results in the breaking of spin-momentum locking. 

Spin in a non-gyromagnetic material emerges due to the presence of out-of-phase field components, which are often introduced due to structural features such as dielectric interfaces or guided-wave structures. For example, it can be shown that a waveguide mode with a longitudinal magnetic field component will have a structure-induced transverse magnetic-spin-profile, which is locked to the direction of momentum. Reversal of momentum will therefore reverse this structure-induced spin-profile. In a waveguiding structure filled with a gyromagnetic material, it might be possible that a conflict arises between the structure-induced momentum-locked spin and the gyrotropy-induced material-locked spin. In the next section, we investigate the nature of wave spin in a guided structure where such a conflict may arise.

\section{Gyrotropy versus spin-momentum locking}

\begin{figure}
    \centering
    \includegraphics[scale=0.7]{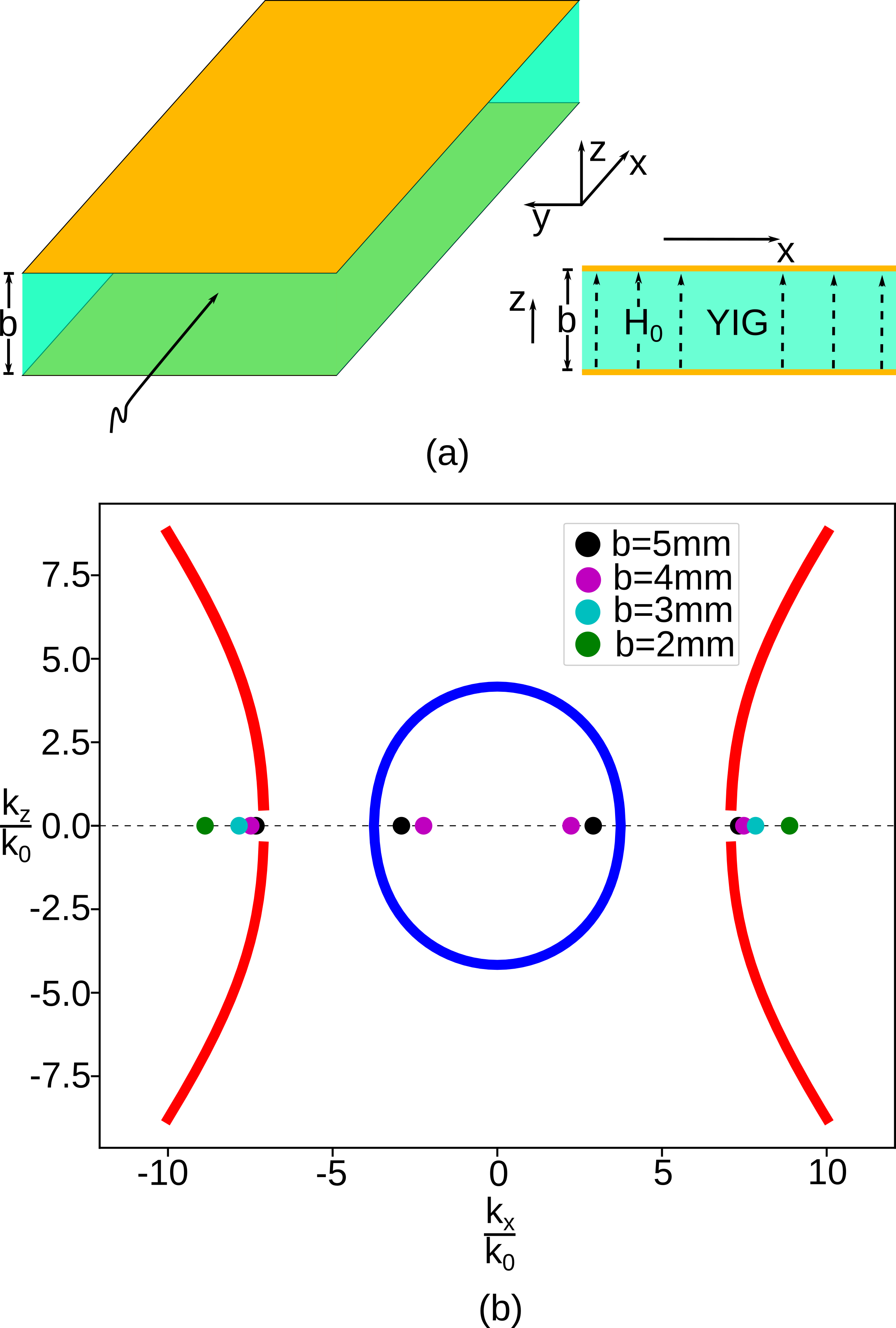}
    \caption{(a) Geometry of the parallel plate waveguide filled with ferrite and having plate separation b, the direction of propagation along the $x-z$ plane,(b) Isofrequency contours representing elliptical and hyperbolic modes for $\mu^\prime=-1.38$ and $\kappa^\prime=2.62$, represented in blue and red color respectively when biased along the $z$-axis,fundamental propagating solutions are shown as dots representing $k_x$ with color representing the different plate separation distance for both modes. }
    \label{fig:sol_mov}
\end{figure}

\begin{figure}
    \centering
    \includegraphics[width=1\linewidth]{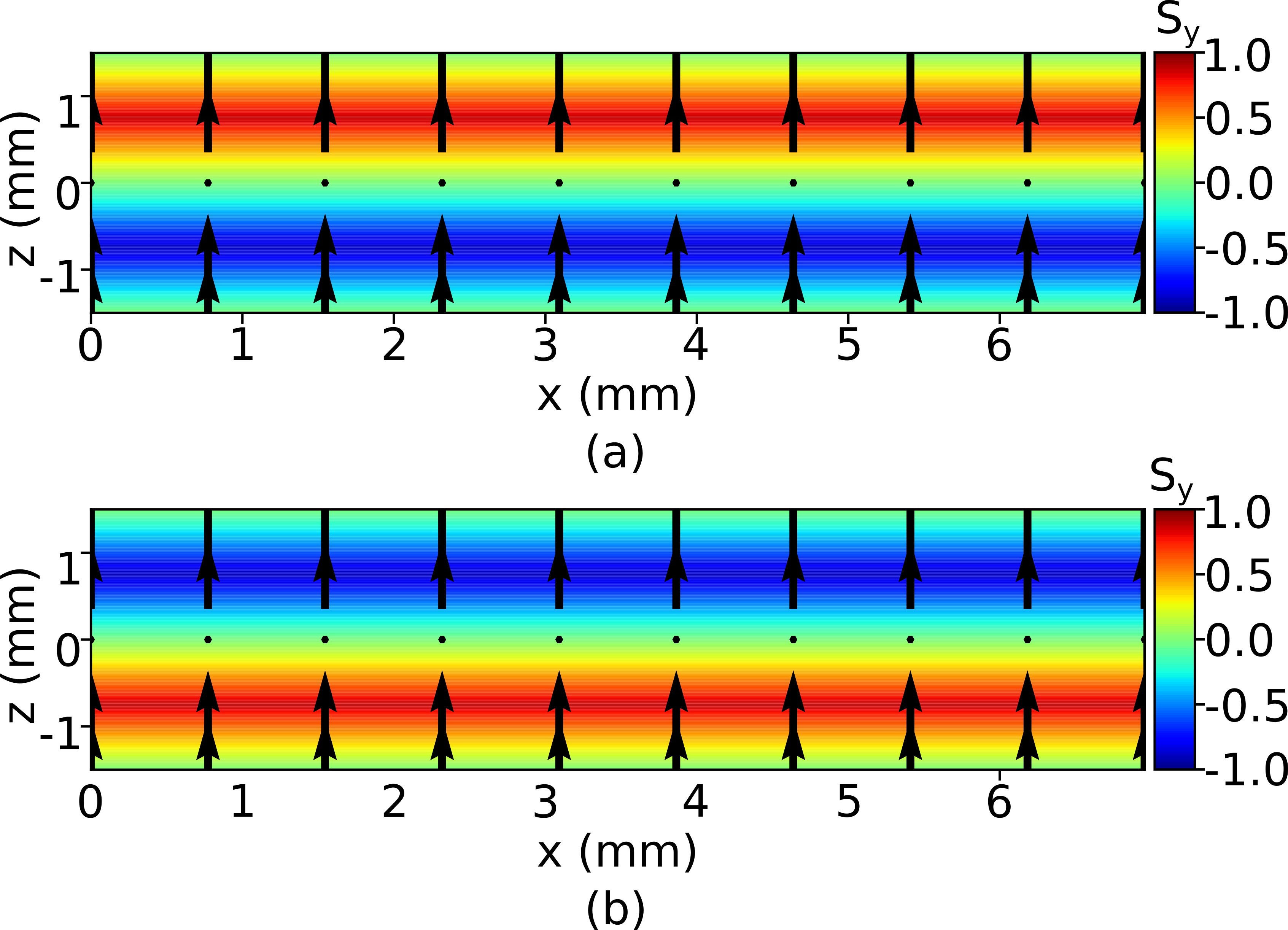}
    \caption{Magnetic field spin-profile along the waveguide. Spin in the $y$-direction is represented through color mapping of the $\hat{S_y}$, and spin in $x-z$ plane is shown as arrows. Panel (a) and (b) show the spin-profile along the waveguide for the lowest order forward and backward propagation, respectively. Plate separation distance $b=3$ mm, $\mu^\prime=-1.38$, and $\kappa^\prime=2.62$.}
    \label{fig:wg_spin}
\end{figure}

To understand the nature of material-induced spin in guided structures, we consider a parallel plate waveguide filled with gyromagnetic material, as shown in Fig.~\ref{fig:sol_mov}(a). We perform a 2D in-plane analysis in the $x-z$ plane, with the direction of propagation being in the $\hat{x}$ direction. Curves in Fig. \ref{fig:sol_mov}(b) show the material topology when the magnetic bias is along the $+\hat{z}$ axis. The dots represent the propagation constant $k_x$ of the waveguide modes for different values of waveguide width $b$. It can be seen that due to the bounded nature of elliptical topology, it supports propagating modes only for waveguide width $b$ greater than a specific value. On the other hand, the hyperbolic mode does not have a cut-off and has a propagating solution even for small values of $b$. Here $k_0$ corresponds to a frequency of 11 GHz. Values of $\mu^\prime$ and $\kappa^\prime$ are -1.38 and 2.62, respectively, and the relative permittivity of the medium $\epsilon_r$ is 14. To solve for the waveguide modes PEC boundary condition of the normal component of magnetic flux density $B_z=0$ is considered along the metallic boundary of the waveguide at $z=-\pm b/2$. Magnetic flux density is calculated from the magnetic field using constitutive relation for gyromagnetic medium, $\Vec{B}=\mu_0\overset{\leftrightarrow}{\mu_r}\Vec{H}$.

Figure~\ref{fig:wg_spin} shows the magnetic spin of the hyperbolic waveguide mode for waveguide width $b=3$~mm.  Magnetic field spin profile for the forward and backward propagating modes are shown in Fig.~\ref{fig:wg_spin}(a) and (b), respectively. Arrows show the spin in the $x-z$ plane, and the color plot represents the spin in the transverse $y$-direction. It can be observed that spin orientation in the $x-z$ plane is solely along the direction of bias ($+\hat{z}$), as shown by arrows. This is the gyrotropy-induced spin and is locked to the medium. Both forward and backward propagating modes have spin along the positive $\hat{z}$ direction. On the other hand, the transverse spin along the $y$-axis is generated due to the guided nature of waveguide mode. This is the structure-induced spin. It can be seen that the structure-induced spin in the $\hat{y}$ direction reverses its profile for forward and backward propagation, implying spin-momentum locking for the transverse spin. Here the material-induced spin is locked to the direction of bias irrespective of the direction of propagation, while the structure-induced spin is locked to the momentum and reverses its profile with the reversal of the direction of propagation. In this case, there is no conflict in the gyrotropy-induced spin and the structure-induced spin.

\begin{figure}
    \centering
    \includegraphics{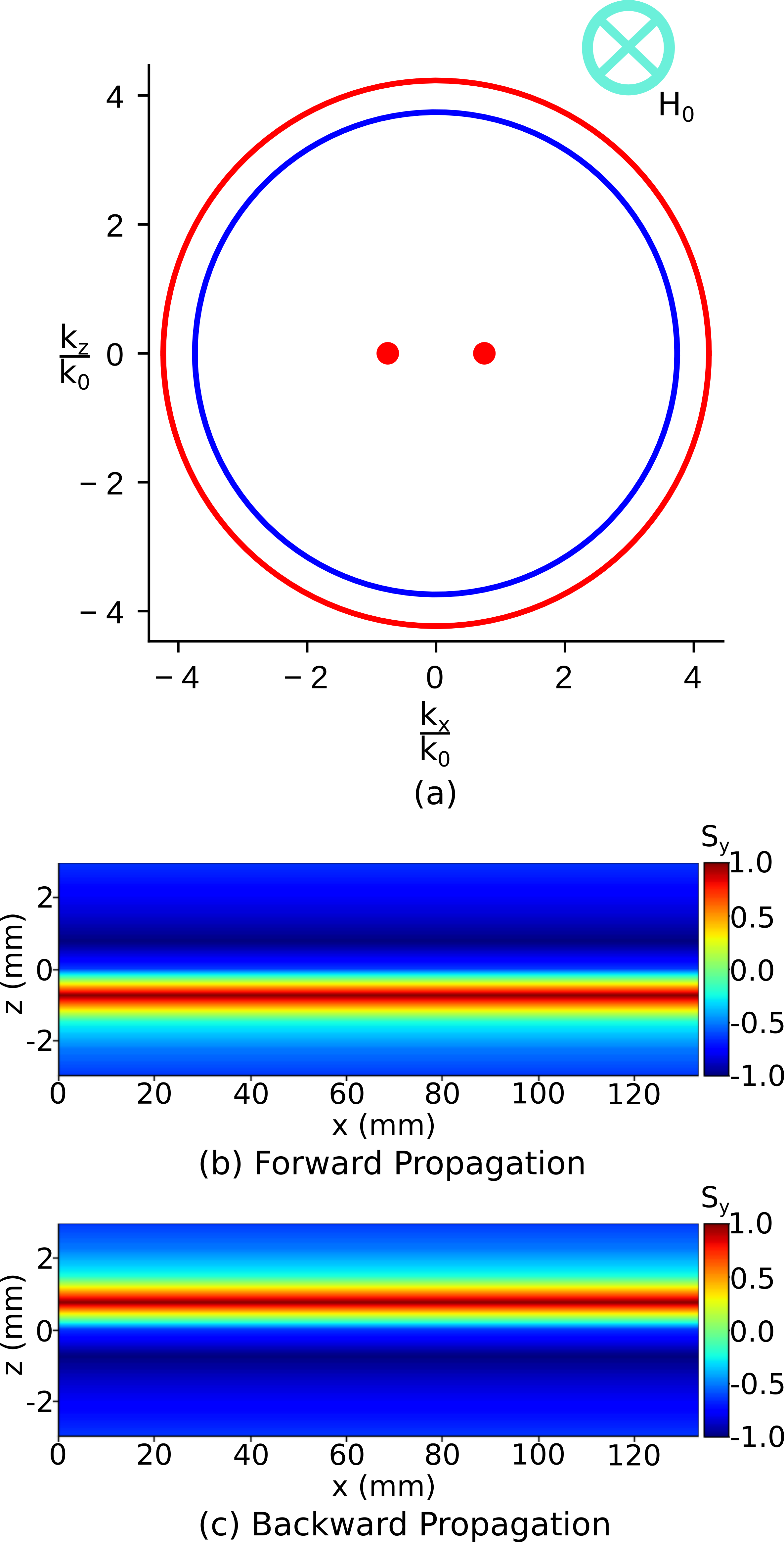}
    \caption{Spin-profile along waveguide for Y-biased ferrite, (a) isofrequency contour for Y-biased structure, dots represent propagating solutions for $k_x$, (b) and (c) show $\hat{S_y}$  spin as a color map. The spin in the $x-z$ plane is zero. The spin tries to align mostly along the direction of material for forward as well as backward propagating modes, thus resulting in an unsymmetric mode profile. Plate separation distance $b=6$ mm, $\mu^\prime=2$ and $\kappa^\prime=1.2$. $k_0$ corresponds to a frequency of 6~GHz.}
    \label{fig:fig9}
\end{figure}

\begin{figure}
    \centering
    \includegraphics{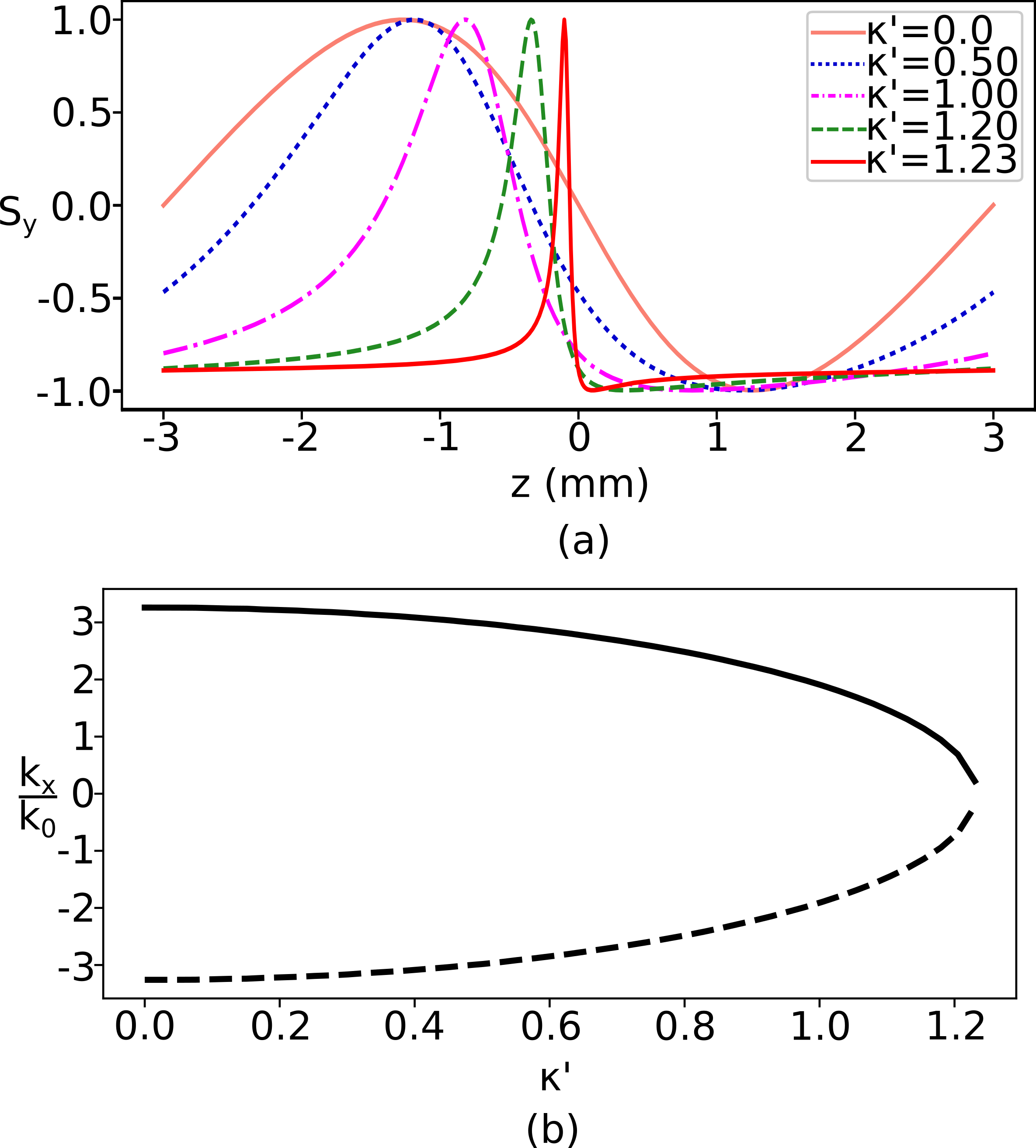}
    \caption{(a) Variation in spin-profile of the waveguide with changing $\kappa^\prime$. As $\kappa^\prime$ increases the mode becomes asymmetric. (b) The shift in value of $k_x$ for fundamental mode with a change in $\kappa^\prime$, $\mu^\prime=2$. Plate separation is 6mm and $k_0$ corresponds to a frequency of 6~GHz. A cut-off in the propagation is observed when gyromagntic term starts dominating the spin profile.}
    \label{fig:kx_mov_and_spins_6g}
\end{figure}

An interesting case arises when the gyrotropy-induced spin and the structure-induced spin are in the same direction. We would expect the material-induced spin to maintain its direction on the reversal of momentum. On the contrary, the structure-induced spin would tend to reverse its direction on account of spin-momentum locking. This happens when the gyromagnetic medium is biased in the $+\hat{y}$, with permeability tensor given by,
\begin{equation}
    \label{eq:y_bias_tensor}
    \overset{\longleftrightarrow}{\mu_{r,y}}=
    \begin{bmatrix}
    \mu^\prime&0&j\kappa^\prime\\
    0&1&0\\
    -j\kappa^\prime&0&\mu^\prime
    \end{bmatrix}
\end{equation}

For $y$-biased material, the 3D topological surface intersects the $k_x-k_z$ plane in two concentric circles (for elliptical as well as hyperbolic regimes), resulting in circular isofrequency curves as shown in Fig.~\ref{fig:fig9}(a). Here $\kappa^\prime = 1.2$ and $\mu^\prime = 2$ is considered with a waveguide width of $b=6mm$, at 6 GHz frequency. The inner circle is a TM mode with zero magnetic spin. However, the modes corresponding to the outer curve have longitudinal ($\hat{x}$ direction) and transverse magnetic ($\hat{z}$ direction) fields, resulting in a transverse spin in the $y$-direction. The forward and backward propagating solutions for this mode are shown by dots in Fig.~\ref{fig:fig9}(a). It is for these modes that a conflict between material-induced and structure-induced spin is expected. The spin-profile for these forward and backward propagating solutions is shown in Fig.~\ref{fig:fig9}(b) and (c), respectively. It can be seen that the spin-profile, in this case, is not symmetric about the center of the waveguide. The negative spin dominates for the forward as well as the backward direction of propagation. Note that the material induced spin, in this case, is in the direction anti-parallel to the direction of magnetic bias, as was discussed in Section-III. With the reversal of direction of propagation, the spin-profile flips about $z=0$. However, the material-induced spin dominates the profiles in both direction of propagation.

As the gyromagnetic term $\kappa^\prime$ increases, the material-induced spin starts dominating the spin-profile, which is otherwise symmetric about the center of the waveguide. This is shown in Fig.~\ref{fig:kx_mov_and_spins_6g}(a) for the forward propagating mode. When $\kappa^\prime=0$, the spin-profile is symmetric. However, as the magnitude of $\kappa^\prime$ increases, the spin-profile starts becoming asymmetric with a greater region of the cross-section trying to align with the material-induced spin. The structure-induced spin, however, tries to maintain the positive and negative spin-profile across the cross-section. As shown in Fig.~\ref{fig:kx_mov_and_spins_6g}(b), this conflict in the gyrotropy-induced spin and the structure-induced spin results in the suppression of propagating waveguide mode. As we increase the magnitude of $\kappa^\prime$, the suppression results in a gyrotropy-induced cut-off of the propagating waveguide solution.

\section{conclusion}

In this paper, we have rigorously investigated topological regimes and spin-profile in a general gyromagnetic medium. We have revealed that the properties as well as the conditions for the existence of isofrequency surfaces are governed by the gyrotropy-induced photonic spin. We have shown that gyrotropy can completely suppress an isofrequency surface or support an otherwise nonexistent surface. Moreover, the spin-profile on topological surfaces depicts a violation of time-reversal symmetry and leads to spin non-reciprocity. Further, we have shown that the material-induced spin violates spin-momentum locking, which results in a conflict with the structure-induced spin. Due to this conflict, an asymmetric mode-profile and a gyrotropy-induced cut-off can be observed in guided wave structures with material gyrotropy. Our work clearly establishes the link between photonic-spin, topological surfaces, and non-reciprocity in a gyromagnetic medium. The results presented in this paper provides important guiding principles for designing materials and structures with engineered topology and photonic-spin. Although we have considered the case of gyromagnetic material and magnetic spin, the insights in this paper will be applicable to gyroelectric media with electric spin as well. Engineering the gyrotropy, along with the permeability and permittivity can provide an additional degree of freedom for designing engineered materials. Further, since gyrotropy can be tuned by the biasing field, the results presented in this paper may lead to a new class of gyrotropy-controlled reconfigurable topological materials. 

\appendix

\section{Derivation of 3d Isofrequency surface and field equations}
\label{ap:iso_field_eq}
Wave propagation in a medium can be understood with the help of its propagation constant $k$ along the principal axes.
These principal components of the propagation constant along $\hat{x}$, $\hat{y}$, and $\hat{z}$ in the polar form are $k_r\sin\theta\cos\phi$, $k_r\sin\theta\sin\phi$, and $k_r\cos\theta$, respectively.
Using these values of $k$ along the principal axes, we find a k-tensor $\overset{\leftrightarrow}{k}$, such that the curl operation for the electric and magnetic field can be replaced with a matrix multiplication with the $\overset{\leftrightarrow}{k}$, i.e., $\nabla\times\vec{E}$ and $\nabla\times\vec{H}$ is equivalent to $\overset{\leftrightarrow}{k}\cdot\vec{E}$ and $\overset{\leftrightarrow}{k}\cdot\vec{H}$, respectively.
The generalized 3D k-tensor $\overset{\leftrightarrow}{k}$ in the polar form is
\begin{equation}
    \label{eq:3d_k_ten_pol}
    \overset{\leftrightarrow}{k}=
    \begin{bmatrix}
    0&-k_r\cos\theta&k_r\sin\theta\sin\phi\\
    k_r\cos\theta&0&-k_r\sin\theta\cos\phi\\
    -k_r\sin\theta\sin\phi&k_r\sin\theta\cos\phi&0\\
    \end{bmatrix}
\end{equation}
Using this $\overset{\leftrightarrow}{k}$ in the wave equation of the form $\det( [\overset{\leftrightarrow}{k}\cdot\overset{\leftrightarrow}{k}+\epsilon_rk_0^2\overset{\leftrightarrow}{\mu_r}])=0$, we get
\begin{multline}
    \label{eq:3d_deter}
    0.5\epsilon_rk_0^2(k_r^4(\mu^\prime+1)+2\epsilon_r^2k_0^4(\mu^{\prime 2}-\kappa^{\prime 2})+\epsilon_rk_0^2k_r^2(\kappa^{\prime 2}-\mu^\prime\\
    (3+\mu^\prime))    -k_r^2(k_r^2(\mu^\prime-1)+\epsilon_rk_0^2(\kappa^{\prime 2}+\mu^\prime-\mu^{\prime 2}))\cos2\theta)=0
\end{multline}
Solving eq.(\ref{eq:3d_deter}) for the roots gives the solution $k_r$.
We get two independent solutions from this biquadratic equation as the two isofrequency surfaces.
\begin{widetext}
\begin{multline}
    \label{eq.3d_isofreq_surface_1}
    k_{r1}=\sqrt{\epsilon_r}k_0((\kappa^{\prime 2}-3\mu^\prime-\mu^{\prime 2}-\cos2\theta(\kappa^{\prime 2}+\mu^\prime-\mu^{\prime 2}) + (8(\mu^{\prime 2}-\kappa^{\prime 2})(-1-\mu^\prime+(\mu^\prime-1)\cos2\theta)\\+(-\kappa^{\prime 2}+\mu^\prime(3+\mu^\prime)+(\kappa^{\prime 2}+\mu^\prime-\mu^{\prime 2})\cos2\theta)^2)^{0.5})/(2(-1-\mu^\prime+(\mu^\prime-1)\cos2\theta)))^{0.5}
\end{multline}
\begin{multline}
    \label{eq.3d_isofreq_surface_2}
    k_{r2}=\sqrt{\epsilon_r}k_0((\kappa^{\prime 2}-3\mu^\prime-\mu^{\prime 2}-\cos2\theta(\kappa^{\prime 2}+\mu^\prime-\mu^{\prime 2}) - (8(\mu^{\prime 2}-\kappa^{\prime 2})(-1-\mu^\prime+(\mu^\prime-1)\cos2\theta)\\+(-\kappa^{\prime 2}+\mu^\prime(3+\mu^\prime)+(\kappa^{\prime 2}+\mu^\prime-\mu^{\prime 2})\cos2\theta)^2)^{0.5})/(2(-1-\mu^\prime+(\mu^\prime-1)\cos2\theta)))^{0.5}
\end{multline}
\end{widetext}
We see that the $k_r$ solutions are independent of the azimuth variable $\phi$, which supplements our approach of analyzing the isofrequency surface as its two dimensional variants as isofrequency contours.
Using the wave equation for the elimination of the magnetic field $[\overset{\leftrightarrow}{k}\cdot\overset{\leftrightarrow}{k}+\epsilon_rk_0^2\overset{\leftrightarrow}{\mu_r}]\cdot\vec{H}=0$, We find the $y$ and $z$ component of the magnetic field with respect to the $x$ component.
\begin{widetext}
\begin{equation}
\label{eq:ap_hy}
    H_y=H_x\frac{j\epsilon_rk_0^2\kappa'\cos\phi+(k_r^2-\epsilon_rk_0^2\mu^\prime)\sin\phi}{(k_r^2-\epsilon_rk_0^2\mu^\prime)\cos\phi-j\epsilon_rk_0^2\kappa^\prime\sin\phi}
\end{equation}
\begin{equation}
\label{eq:ap_hz}
    H_z=H_x\frac{(k_r^4-3\epsilon_rk_0^2k_r^2\mu^\prime+2\epsilon_r^2k_0^4(\mu^{\prime 2}-\kappa^{\prime 2})+k_r^2(k_r^2-\epsilon_rk_0^2\mu^\prime)\cos2\theta)\csc\theta\sec\theta}{2k_r^2((k_r^2-\epsilon_rk_0^2\mu^\prime)\cos\phi-j\epsilon_rk_0^2\kappa^\prime\sin\phi)}
\end{equation}
\end{widetext}
Equation \ref{eq:ap_hy} and \ref{eq:ap_hz} gives the $H_y$ and $H_z$ component of the magnetic field with respect to $H_x$, respectively.

\section{Definition of magnetic spin}
The third Stoke's parameter $S_3$ denotes the spin sense and its magnitude in an arbitrary direction.
The magnetic field vector $\vec{H}$ in the Cartesian coordinate system can be written in terms of its directional elements as
\begin{equation}
    \label{eq:h_vec_norm}
    \vec{H}=H_x\hat{x}+H_y\hat{y}+H_z\hat{z}
\end{equation}
Similarly, the conjugate magnetic field vector $\vec{H}^*$ is
\begin{equation}
    \label{eq:h_vec_conj}
    \vec{H}^*=H_x^*\hat{x}+H_y^*\hat{y}+H_z^*\hat{z}
\end{equation}
The individual directional components of $\vec{H}$ are
\begin{equation}
    \label{eq:h_vec_comps}
    \begin{split}
        H_x=|H_x|\exp{j\phi_x}\\
        H_y=|H_y|\exp{j\phi_y}\\
        H_z=|H_z|\exp{j\phi_z}\\
    \end{split}
\end{equation}
Cross product of $\vec{H}$ and $\vec{H}^*$ is
\begin{multline}
    \label{eq:mag_cross_prod}
    \vec{H}^*\times\vec{H}=2j|H_y||H_z|\sin(\phi_z-\phi_y)\hat{x} + 2j|H_z||H_x|\\ \sin(\phi_x-\phi_z)\hat{y} + 2j|H_x||H_y|\sin(\phi_y-\phi_x)\hat{z}
\end{multline}
The imaginary part of eq.(\ref{eq:mag_cross_prod}) is
\begin{equation}
    \label{eq:h_crprod_in_s3}
    Im(\vec{H}^*\times\vec{H})=S_{3x}\hat{x}+S_{3y}\hat{y}+S_{3z}\hat{z}
\end{equation}
The third Stokes parameter $S_3$ is sufficient to describe the spin-profile in any arbitrary direction as a vector sum of spins along the principal axes.
The third Stokes parameter for the principal axes are
\begin{equation}
    \begin{split}
        S_{3x}=2|H_y||H_z|\sin(\phi_z-\phi_y)\\
        S_{3y}=2|H_z||H_x|\sin(\phi_x-\phi_z)\\
        S_{3z}=2|H_x||H_y|\sin(\phi_y-\phi_x)\\
    \end{split}
\end{equation}

\section{Derivation of the spin-crossover angle}
Spin-crossover angle acts as separating the angular region, defining parallel and anti-parallel spin. 
We use a simplified polar form for defining this spin-crossover angle, which requires the measurement of angle to start from $k_x$ axis and traverse counter-clockwise.
This leads to a modified $\overset{\leftrightarrow}{k}_{sc}$ tensor in the Cartesian form
\begin{equation}
    \overset{\leftrightarrow}{k}_{sc}=
    \begin{bmatrix}
    0&-k_r\sin(\zeta_{sc})&0\\
    k_r\sin(\zeta_{sc})&0&-k_r\cos(\zeta_{sc})\\
    0&k_r\cos(\zeta_{sc})&0\\
    \end{bmatrix}
\end{equation}
Using modified wave equation $ [\overset{\leftrightarrow}{k}_{sc}\cdot\overset{\leftrightarrow}{k}_{sc}+\epsilon_rk_0^2\overset{\leftrightarrow}{\mu_r}]\cdot[\vec{H}]=0$, the matrix defining the field component relationship is derived. 
Magnetic field spin in the $x-y$ plane is defined as
\begin{equation}
    \label{eq:fld_sc_hy}
    H_x=-H_y\frac{\epsilon_rk_0^2\mu^\prime-k_r^2}{j\epsilon_rk_0^2\kappa^\prime}
\end{equation}
We observe $k_r=k_0\sqrt{\epsilon_r\mu^\prime}$ leads to zero spin along the bias axis. 
This point is termed as the spin-crossover point as it acts as the boundary separating parallel and anti-parallel material spin.
Solving the modified wave equation for $k_r$, we get two expressions representing the two independent isofrequency surfaces.
Only one of these isofrequency surfaces demonstrates spin-inversion corresponding to the angle of propagation.
Equating this expression of $k_r$ with $k_0\sqrt{\epsilon_r\mu^\prime}$, we get the spin-crossover angle as
\begin{equation}
    \label{eq:spin_sc_eq}
    \zeta_{sc}=\cos^{-1}\left(\sqrt{\frac{1}{\mu^\prime}}\right)
\end{equation}

%appendixname

% \bibliographystyle{plain}\textsl{}
\bibliographystyle{ieeetr}
\renewcommand \bibname{References}
\bibliography{reference.bib}

\begin{thebibliography}{10}

\bibitem{non_recp_fer_2020_1}
H.~Ma, C.~Ju, X.~Xi, and R.-X. Wu, ``Nonreciprocal goos-hanchen shift by
  topological edge states of a magnetic photonic crystal,'' {\em Opt. Express},
  pp.~19916--19925, Jul 2020.

\bibitem{non_recp_fer_2020_2}
G.~Portela, V.~Dmitriev, and D.~Zimmer, ``Ferromagnetic resonance isolator
  based on a photonic crystal structure with terahertz vortices,'' {\em
  Photonic Network Communications}, vol.~39, pp.~47--53, Feb 2020.

\bibitem{non_recp_fer_2018_1}
N.~{Parsa} and R.~C. {Toonen}, ``Ferromagnetic nanowires for nonreciprocal
  millimeter-wave applications: Investigations of artificial ferrites for
  realizing high-frequency communication components,'' {\em IEEE Nanotechnology
  Magazine}, vol.~12, no.~4, pp.~28--35, 2018.

\bibitem{non_recp_fer_2019_1}
A.~Savchenko and V.~Krivoruchko, ``Electric-field control of nonreciprocity of
  spin wave excitation in ferromagnetic nanostripes,'' {\em Journal of
  Magnetism and Magnetic Materials}, vol.~474, pp.~9 -- 13, 2019.

\bibitem{non_rec_meta_surf}
D.~Frese, Q.~Wei, Y.~Wang, L.~Huang, and T.~Zentgraf, ``Nonreciprocal
  asymmetric polarization encryption by layered plasmonic metasurfaces,'' {\em
  Nano Letters}, vol.~19, pp.~3976--3980, Jun 2019.

\bibitem{nnrcproc_ncom_gnt}
D.~L. Sounas, C.~Caloz, and A.~Al{\`u}, ``Giant non-reciprocity at the
  subwavelength scale using angular momentum-biased metamaterials,'' {\em
  Nature Communications}, vol.~4, p.~2407, Sep 2013.

\bibitem{anti_fer_mag_hyper}
R.~Mac\^edo and R.~E. Camley, ``Engineering terahertz surface magnon-polaritons
  in hyperbolic antiferromagnets,'' {\em Phys. Rev. B}, vol.~99, p.~014437, Jan
  2019.

\bibitem{chern_insul_2020_1}
R.-C. Shiu, H.-C. Chan, H.-X. Wang, and G.-Y. Guo, ``Photonic chern insulators
  made of gyromagnetic hyperbolic metamaterials,'' {\em Phys. Rev. Materials},
  vol.~4, p.~065202, Jun 2020.

\bibitem{dm_pozzar_book}
D.~M. Pozar, {\em Microwave engineering}.
\newblock Wiley; Fourth edition, India., 2012.

\bibitem{spin_one_way_gyromag}
Z.~Li and R.-x. Wu, ``An experimental study of self-guided unidirectional
  waveguides by a chain of gyro-magnetic rods,'' {\em Applied Physics A},
  vol.~124, p.~139, Jan 2018.

\bibitem{Lokk2017_gyromag_iso_freq_surface}
E.~G. Lokk, ``Isofrequency surfaces and dependences of electromagnetic waves in
  infinite ferromagnetic space,'' {\em Journal of Communications Technology and
  Electronics}, vol.~62, pp.~251--259, Mar 2017.

\bibitem{arnaud2020_ferrite_ant_circ_patch_LEO_sat}
E.~Arnaud, L.~Huitema, R.~Chantalat, A.~Bellion, and T.~Monediere, ``Circularly
  polarized ferrite patch antenna for leo satellite applications,'' {\em
  International Journal of Microwave and Wireless Technologies}, vol.~12,
  no.~4, p.~332–338, 2020.

\bibitem{siw_mod_conv_2019}
A.~{Afshani} and K.~{Wu}, ``Nonreciprocal mode converting waveguide and
  circulator,'' {\em IEEE Transactions on Microwave Theory and Techniques},
  vol.~67, no.~8, pp.~3350--3360, 2019.

\bibitem{9238458}
Y.~{Zhang}, D.~{Cai}, C.~{Zhao}, M.~{Zhu}, Y.~{Gao}, Y.~{Chen}, X.~{Liang},
  H.~{Chen}, J.~{Wang}, Y.~{Wei}, Y.~{He}, C.~{Dong}, N.~{Sun},
  M.~{Zaeimbashi}, Y.~{Yang}, H.~{Zhu}, B.~{Zhang}, K.~{Huang}, and N.~X.
  {Sun}, ``Nonreciprocal isolating bandpass filter with enhanced isolation
  using metallized ferrite,'' {\em IEEE Transactions on Microwave Theory and
  Techniques}, vol.~68, no.~12, pp.~5307--5316, 2020.

\bibitem{nano_conduits_magnonics_2020}
B.~Heinz, T.~Brächer, M.~Schneider, Q.~Wang, B.~Lägel, A.~M. Friedel,
  D.~Breitbach, S.~Steinert, T.~Meyer, M.~Kewenig, C.~Dubs, P.~Pirro, and A.~V.
  Chumak, ``Propagation of spin-wave packets in individual nanosized yttrium
  iron garnet magnonic conduits,'' {\em Nano Letters}, vol.~20, no.~6,
  pp.~4220--4227, 2020.
\newblock PMID: 32329620.

\bibitem{stancil2009spin_book}
D.~D. Stancil and A.~Prabhakar, {\em Spin waves}, vol.~5.
\newblock Springer, 2009.

\bibitem{chap_2_book_magnon}
R.~Macêdo, ``Chapter two - tunable hyperbolic media: Magnon-polaritons in
  canted antiferromagnets,'' vol.~68 of {\em Solid State Physics}, pp.~91 --
  155, Academic Press, 2017.

\bibitem{spin_polariton_two_dim}
C.~E. Whittaker, E.~Cancellieri, P.~M. Walker, D.~R. Gulevich, H.~Schomerus,
  D.~Vaitiekus, B.~Royall, D.~M. Whittaker, E.~Clarke, I.~V. Iorsh, I.~A.
  Shelykh, M.~S. Skolnick, and D.~N. Krizhanovskii, ``Exciton polaritons in a
  two-dimensional lieb lattice with spin-orbit coupling,'' {\em Phys. Rev.
  Lett.}, vol.~120, p.~097401, Mar 2018.

\bibitem{magnon_photon_phonon_ent}
J.~Li, S.-Y. Zhu, and G.~S. Agarwal, ``Magnon-photon-phonon entanglement in
  cavity magnomechanics,'' {\em Phys. Rev. Lett.}, vol.~121, p.~203601, Nov
  2018.

\bibitem{magnon_photon_coupling}
M.~Harder, Y.~Yang, B.~M. Yao, C.~H. Yu, J.~W. Rao, Y.~S. Gui, R.~L. Stamps,
  and C.-M. Hu, ``Level attraction due to dissipative magnon-photon coupling,''
  {\em Phys. Rev. Lett.}, vol.~121, p.~137203, Sep 2018.

\bibitem{phot_spin_2012_1}
K.~Y. Bliokh and F.~Nori, ``Transverse spin of a surface polariton,'' {\em
  Physical review A}, vol.~85, no.~6, p.~061801, 2012.

\bibitem{todd_spin_lock}
T.~V. Mechelen and Z.~Jacob, ``Universal spin-momentum locking of evanescent
  waves,'' {\em Optica}, vol.~3, pp.~118--126, Feb 2016.

\bibitem{spin_mom_lock_cuprate}
K.~Gotlieb, C.-Y. Lin, M.~Serbyn, W.~Zhang, C.~L. Smallwood, C.~Jozwiak,
  H.~Eisaki, Z.~Hussain, A.~Vishwanath, and A.~Lanzara, ``Revealing hidden
  spin-momentum locking in a high-temperature cuprate superconductor,'' {\em
  Science}, vol.~362, no.~6420, pp.~1271--1275, 2018.

\bibitem{spin_mom_lock_nature}
S.~Luo, L.~He, and M.~Li, ``Spin-momentum locked interaction between guided
  photons and surface electrons in topological insulators,'' {\em Nature
  Communications}, vol.~8, p.~2141, Dec 2017.

\bibitem{farid_spin_lock}
F.~Kalhor, T.~Thundat, and Z.~Jacob, ``Universal spin-momentum locked optical
  forces,'' {\em Applied Physics Letters}, vol.~108, no.~6, p.~061102, 2016.

\bibitem{spin_opt_forces_trans_nat}
H.~Magallanes and E.~Brasselet, ``Macroscopic direct observation of optical
  spin-dependent lateral forces and left-handed torques,'' {\em Nature
  Photonics}, vol.~12, pp.~461--464, Aug 2018.

\bibitem{spin_opt_forces_opti_trap}
V.~Svak, O.~Brzobohat{\'y}, M.~{\v{S}}iler, P.~J{\'a}kl, J.~Ka{\v{n}}ka,
  P.~Zem{\'a}nek, and S.~H. Simpson, ``Transverse spin forces and
  non-equilibrium particle dynamics in a circularly polarized vacuum optical
  trap,'' {\em Nature Communications}, vol.~9, p.~5453, Dec 2018.

\bibitem{hyp_meta_1}
H.~N.~S. Krishnamoorthy, Y.~Zhou, S.~Ramanathan, E.~Narimanov, and V.~M. Menon,
  ``Tunable hyperbolic metamaterials utilizing phase change heterostructures,''
  {\em Applied Physics Letters}, vol.~104, no.~12, p.~121101, 2014.

\bibitem{hyp_meta_2}
V.~R. Tuz, I.~V. Fedorin, and V.~I. Fesenko, ``Bi-hyperbolic isofrequency
  surface in a magnetic-semiconductor superlattice,'' {\em Opt. Lett.},
  vol.~42, pp.~4561--4564, Nov 2017.

\bibitem{hyp_meta_3}
A.~Poddubny, I.~Iorsh, P.~Belov, and Y.~Kivshar, ``Hyperbolic metamaterials,''
  {\em Nature Photonics}, vol.~7, pp.~948--957, Dec 2013.

\bibitem{hyp_meta_4}
M.~Durach, ``Tetra-hyperbolic and tri-hyperbolic optical phases in anisotropic
  metamaterials without magnetoelectric coupling due to hybridization of
  plasmonic and magnetic bloch high-k polaritons,'' {\em Optics
  Communications}, vol.~476, p.~126349, 2020.

\bibitem{hyp_meta_5}
P.~Shekhar, J.~Atkinson, and Z.~Jacob, ``Hyperbolic metamaterials: fundamentals
  and applications,'' {\em Nano Convergence}, vol.~1, p.~14, Jun 2014.

\bibitem{hyper_6}
X.~Song, Z.~Liu, Y.~Xiang, and K.~Aydin, ``Biaxial hyperbolic metamaterials
  using anisotropic few-layer black phosphorus,'' {\em Opt. Express}, vol.~26,
  pp.~5469--5477, Mar 2018.

\bibitem{hyperbolic_hyper_lens}
Z.~Jacob, L.~V. Alekseyev, and E.~Narimanov, ``Optical hyperlens: Far-field
  imaging beyond the diffraction limit,'' {\em Opt. Express}, vol.~14,
  pp.~8247--8256, Sep 2006.

\bibitem{super_lense_near_field}
A.~Ghoshroy, W.~Adams, X.~Zhang, and D.~O. G\"{u}ney, ``Enhanced superlens
  imaging with loss-compensating hyperbolic near-field spatial filter,'' {\em
  Opt. Lett.}, vol.~43, pp.~1810--1813, Apr 2018.

\bibitem{therm_rad_hyp}
J.~Shi, B.~Liu, P.~Li, L.~Y. Ng, and S.~Shen, ``Near-field energy extraction
  with hyperbolic metamaterials,'' {\em Nano Letters}, vol.~15, pp.~1217--1221,
  Feb 2015.

\bibitem{liu2017integrated}
F.~Liu, L.~Xiao, Y.~Ye, M.~Wang, K.~Cui, X.~Feng, W.~Zhang, and Y.~Huang,
  ``Integrated cherenkov radiation emitter eliminating the electron velocity
  threshold,'' {\em Nature Photonics}, vol.~11, no.~5, p.~289, 2017.

\bibitem{cherenkov_natur_hyper_metarial}
J.~Tao, L.~Wu, G.~Zheng, and S.~Yu, ``Cherenkov polaritonic radiation in a
  natural hyperbolic material,'' {\em Carbon}, vol.~150, pp.~136 -- 141, 2019.

\bibitem{hex_boron_nitride}
S.~Dai, Q.~Ma, M.~K. Liu, T.~Andersen, Z.~Fei, M.~D. Goldflam, M.~Wagner,
  K.~Watanabe, T.~Taniguchi, M.~Thiemens, F.~Keilmann, G.~C. A.~M. Janssen,
  S.-E. Zhu, P.~Jarillo-Herrero, M.~M. Fogler, and D.~N. Basov, ``Graphene on
  hexagonal boron nitride as a tunable hyperbolic metamaterial,'' {\em Nature
  Nanotechnology}, vol.~10, pp.~682--686, Aug 2015.

\bibitem{hex_boron_nitride_nat}
J.~D. Caldwell, I.~Aharonovich, G.~Cassabois, J.~H. Edgar, B.~Gil, and D.~N.
  Basov, ``Photonics with hexagonal boron nitride,'' {\em Nature Reviews
  Materials}, vol.~4, pp.~552--567, Aug 2019.

\bibitem{neg_ref_fer_2015_1}
C.~Lan, K.~Bi, J.~Zhou, and B.~Li, ``Experimental demonstration of hyperbolic
  property in conventional material—ferrite,'' {\em Applied Physics Letters},
  vol.~107, no.~21, p.~211112, 2015.

\bibitem{CHENG20183018}
Y.~Cheng, B.~Peng, Z.~Hu, Z.~Zhou, and M.~Liu, ``Recent development and status
  of magnetoelectric materials and devices,'' {\em Physics Letters A},
  vol.~382, no.~41, pp.~3018 -- 3025, 2018.

\bibitem{Pendharker:18}
S.~Pendharker, F.~Kalhor, T.~V. Mechelen, S.~Jahani, N.~Nazemifard, T.~Thundat,
  and Z.~Jacob, ``Spin photonic forces in non-reciprocal waveguides,'' {\em
  Opt. Express}, vol.~26, pp.~23898--23910, Sep 2018.

\end{thebibliography}
\end{document}